\pdfoutput=1

\documentclass[11pt]{article}

\usepackage[preprint]{acl}

\usepackage{enumitem}
\usepackage{times}
\usepackage{latexsym}
\usepackage{amsmath}
\usepackage{amssymb}
\usepackage{multirow}
\usepackage{booktabs}
\usepackage{graphicx}
\usepackage{subcaption}
\usepackage{hyperref}
\usepackage{adjustbox}
\usepackage[T1]{fontenc}

\usepackage[utf8]{inputenc}

\usepackage{microtype}

\usepackage{inconsolata}
\usepackage[most]{tcolorbox}
\usepackage{listings}

\usepackage{graphicx}
\usepackage{amsthm}
\theoremstyle{definition}
\newtheorem{definition}{Definition}
\usepackage{tablefootnote}
\usepackage[normalem]{ulem}\usepackage{listings}
\usepackage{xcolor}
\usepackage{colortbl}
\definecolor{codegreen}{rgb}{0,0.6,0}
\definecolor{codegray}{rgb}{0.5,0.5,0.5}
\definecolor{codepurple}{rgb}{0.58,0,0.82}
\definecolor{backcolour}{rgb}{0.95,0.95,0.92}

\lstdefinestyle{mystyle}{
  label=code:sample,
  frame=single,
  backgroundcolor=\color{backcolour}, commentstyle=\color{codegreen},
  keywordstyle=\color{magenta},
  numberstyle=\tiny\color{codegray},
  stringstyle=\color{codepurple},
  basicstyle=\ttfamily\footnotesize,
  breakatwhitespace=false,         
  breaklines=true,                 
  captionpos=t,                    
  keepspaces=true,                 
  numbers=left,                    
  numbersep=5pt,                  
  showspaces=false,                
  showstringspaces=false,
  showtabs=false,                  
  tabsize=1
}

\newtcolorbox{llmpromptbox}[2][]{%
  colback=blue!5,         
  colframe=blue!75,        
  fonttitle=\bfseries,     
  title={#2},             
  fontupper=\small,
  sharp corners,          
  #1                      
}

\title{\textit{Can LLMs Reason About Program Semantics?}\par A Comprehensive Evaluation of LLMs on Formal Specification Inference}

\author{Thanh Le-Cong, Bach Le, Toby Murray \\
        School of Computing and Information Systems \\
        The University of Melbourne \\
        \texttt{congthanh.le@student.unimelb.edu.au}, \texttt{\{bach.le, toby.murray\}@unimelb.edu.au}}
\newboolean{showcomments}
\setboolean{showcomments}{true}
\ifthenelse{\boolean{showcomments}}
{ }

\begin{document}
\maketitle

\begin{abstract}
Large Language Models (LLMs) are increasingly being used to automate programming tasks.
However, the capabilities of LLMs in reasoning about program semantics are still inadequately studied, leaving substantial potential for further exploration.
This paper introduces FormalBench, a comprehensive benchmark designed to evaluate the reasoning abilities of Large Language Models (LLMs) on program semantics. Specifically, it utilizes the task of synthesizing formal program specifications as a proxy measure for assessing the semantic reasoning of LLMs. This task requires both comprehensive reasoning over all possible program executions and the generation of precise, syntactically correct expressions that adhere to formal syntax and semantics. Using this benchmark, we evaluated the ability of LLMs to synthesize consistent and complete specifications. Our findings show that LLMs perform well with simple control flows but struggle with more complex structures, especially loops, even with advanced prompting. Additionally, LLMs exhibit limited robustness against semantic-preserving transformations. We also highlight common failure patterns and design self-repair prompts, improving success rates by 25\%.
FormalBench is packaged as an executable library and has been released at \href{link}{https://github.com/thanhlecongg/FormalBench/}.
\end{abstract}

\section{Introduction}~\label{sec:intro}

Recent advances in Large Language Models (LLMs) have demonstrated substantial potential for code understanding and generation~\cite{hou2024large, chen2021evaluating}. However, as adoption grows, critical concerns emerge about their reliability in programming tasks, particularly their capacity to reason about program semantics~\cite{liu2024your, yang2024robustness, liu2024refining}. A fundamental question remains: \textit{Can LLMs reason about program semantics?} 
Pioneering studies~\cite{pei2023can, chen2024reasoning} have tackled this challenge by evaluating LLMs on \textit{partial} semantic properties, such as predicting execution traces or inferring likely program invariants. Although these efforts provide valuable insights, they examine narrow aspects of program behavior rather than a comprehensive semantic understanding. For example, execution-based evaluations~\cite{chen2024reasoning, jain2024livecodebench} are limited to specific execution paths and inputs, offering an incomplete view of LLMs' semantic reasoning.

\begin{figure}[t]
  \centering
  \begin{lstlisting}[aboveskip=0pt, belowskip=0pt, basicstyle=\scriptsize\ttfamily]
//@ requires num >= 0 && t >= 0;
//@ requires num + 2*t <= Integer.MAX_VALUE;
//@ requires num + 2*t >= Integer.MIN_VALUE;
//@ ensures \result == num + 2*t;
public int theMaximumAchievableX(int num, int t) {
  int res = num;
  //@ maintaining res == num + 2*(i-1);
  //@ maintaining i >= 1 && i <= t+1;
  //@ decreasing t-i+1;
  for(int i = 1; i <= t; i++) {
    res = res + 2;
  }
  return res;
}
  \end{lstlisting}
  \caption{Illustration of a Java program annotated with JML specifications (highlighted in \color{green!50!black}{green}).}
  \label{fig:example}
  \vspace{-5mm}
\end{figure}


Recent studies~\cite{wen2024enchanting, ma2024specgen} have evaluated LLMs in the synthesis of program specifications expressed in formal languages such as JML~\cite{leavens2006preliminary} and ACSL~\cite{baudin2008acsl}. The synthesized specifications can then be used to assist in automated software verification~\cite{d2008survey} and bug finding~\cite{le2022finding}. This task challenges LLMs to (1) reason exhaustively over all possible program executions and (2) generate logically precise expressions that comply with the formal syntax and semantics of the specification language.

While initial results are promising, current evaluation methodologies for LLM reasoning through formal specification inference face three key limitations. 
First, the evaluation datasets are \textit{small and lack diversity}. For example, SpecGenBench~\cite{ma2024specgen} and the Frama-C problems~\cite{kirchner2015frama} contain only 120 and 57 programs, respectively. 
Second, evaluation metrics focus \textit{narrowly on consistency}, i.e., alignment between specifications and programs, while neglecting completeness, i.e., coverage of all semantic behaviors. Completeness is particularly important for assessing LLM's ability to reason comprehensively about \textit{complete} program behaviors.
Finally, current studies primarily aim to develop new LLM-based techniques for formal specification inference rather than \textit{evaluating the LLM reasoning capabilities} themselves. Consequently, evaluations are often ad hoc, relying on specific prompting techniques or LLMs, leading to a lack of comprehensive insights across a wide range of models and prompts.

To address the above challenges, we introduce FormalBench, a comprehensive benchmark for evaluating the reasoning capabilities of LLMs through formal specification inference. FormalBench improves the existing dataset with two notable features: (1) a large-scale dataset of 700 manually validated Java programs and 6,219 augmented programs, covering various control flow structures, and (2) a Python library with a robust suite of evaluation metrics to measure both consistency (via deductive verification) and completeness (through mutation analysis). We then leverage FormalBench to conduct a comprehensive study evaluating eight state-of-the-art LLMs across four critical dimensions:  (1) their effectiveness in synthesizing complete and consistent specifications, (2) robustness against semantic-preserving code transformations, (3) the impact of advanced prompting techniques, and (4) root causes of failures and their self-repair ability. 
 
 Our findings reveal several key insights. LLMs demonstrate limited effectiveness, achieving only about 10\% verification success with over 50\% failures, particularly struggling with complex control-flow structures such as nested loops. Advanced prompting techniques, such as few-shot and least-to-most prompting, improve success rates to 16.6\% and reduce failures; yet, overall performance remains suboptimal. Robustness issues also arise, with LLMs exhibiting flip rates between 27.2\% and 39.2\% under semantic-preserving transformations, negatively impacting their performance. Common failures of LLMs include syntax errors, flawed inductive reasoning, incorrect postconditions, faulty loop invariants, and misjudged arithmetic bounds. However, error-specific prompts enhance LLM self-repair capabilities, improving verifiable specifications by approximately 25\% and reducing failures by around 40\%; although these improvements converge after a few iterations.

In summary, our main contributions include:
\begin{itemize}[noitemsep, left=1pt]
    \item We introduce FormalBench, a comprehensive dataset and toolset designed for evaluating formal reasoning of LLMs about program semantics.
    \item We propose a robust set of evaluation metrics to assess the effectiveness LLMs on synthesizing consistent and complete formal specifications.
    \item We conduct an extensive empirical study of popular LLMs using FormalBench, highlighting their limitations. We also identify common failure patterns and design customized prompts to assist LLMs in self-repairing these failures.
    \item We advance research in formal specification inference by releasing FormalBench as an installable Python library under Apache 2.0 License, lowering barriers for academic research and establishing foundational benchmarks and metrics for future work.
\end{itemize}

\section{Problem Statement}~\label{sec:statement}
Given an input program, the formal specification inference task is to annotate the program with a set of formal specifications, i.e., Boolean expressions written in a formal specification language.
A good formal specification should be adequate, consistent, unambiguous, complete, satisfiable, and minimal~\cite{lamsweerde2000formal}. In this work, we particularly focus on two key properties of specifications, namely completeness and consistency, which reflect the correctness of the generated specifications. Inspired by ~\cite{lamsweerde2000formal}, we define these properties in our problem as follows:

\begin{definition}
(Consistency) A formal specification is considered \textbf{consistent} to be a given input program if all specified properties are well-formed and true with respect to that program.
\end{definition}

\begin{definition}
(Completeness) A formal specification is considered \textbf{complete} if all function properties that hold with respect to a given input program are specified in the specification.
\end{definition}

To determine the consistency of LLM-generated specifications, i.e., specifications that hold for an input program, we use deductive verification tools that transform the annotated program into logical proof obligations and verify them with theorem provers. These tools ensure software correctness by systematically analyzing all possible execution paths, making our consistency checking reliable.  

Measuring the completeness of formal specifications is inherently challenging because of the complex behaviors exhibited by software programs. 
Inspired by the success of mutation testing~\cite{andrews2005mutation} in evaluating the completeness of test suites, we propose to use mutation analysis as a proxy to assess the completeness of formal specifications. Mutation analysis generates non-equivalent mutant variants of input programs by introducing artificial faults~\cite{andrews2005mutation}. Ideally, a complete specification should be able to detect all such faults. Therefore, we measure the proportion of mutants that violate the specification as a proxy of its completeness.

\section{Dataset Construction and Evaluation}

\subsection{FormalBench Construction}~\label{sec:benchmark}

FormalBench is constructed in three phases to ensure its reliability and diversity: (1) curating reference Java programs paired with natural language descriptions, (2) manually verifying program correctness with respect to natural language descriptions to establish FormalBench-Base, which comprises 700 programs, and (3) augmenting FormalBench-Base using semantic-preserving transformations to create FormalBench-Diverse, which comprises 6,219 programs.

Specifically, we begin with an initial pool of 966 Java programs generated by the MBXP model~\cite{athiwaratkun2022multi} for the MBJP benchmark. This dataset is released under the Apache License 2.0, which permits modification and redistribution.
To filter incorrect candidates, we execute these programs against the MBJP test suite, retaining 824 programs that pass all the provided test cases. However, as MBJP’s test suite is generally weak, we conduct manual validation to ensure alignment between program behavior and natural-language intents. This involves a multi-step review process: we carefully inspect each program, augment the test suite with adversarial inputs, and validate correctness against the specified intents. 

\begin{figure}[t]
  \centering
  \includegraphics[width=\columnwidth]{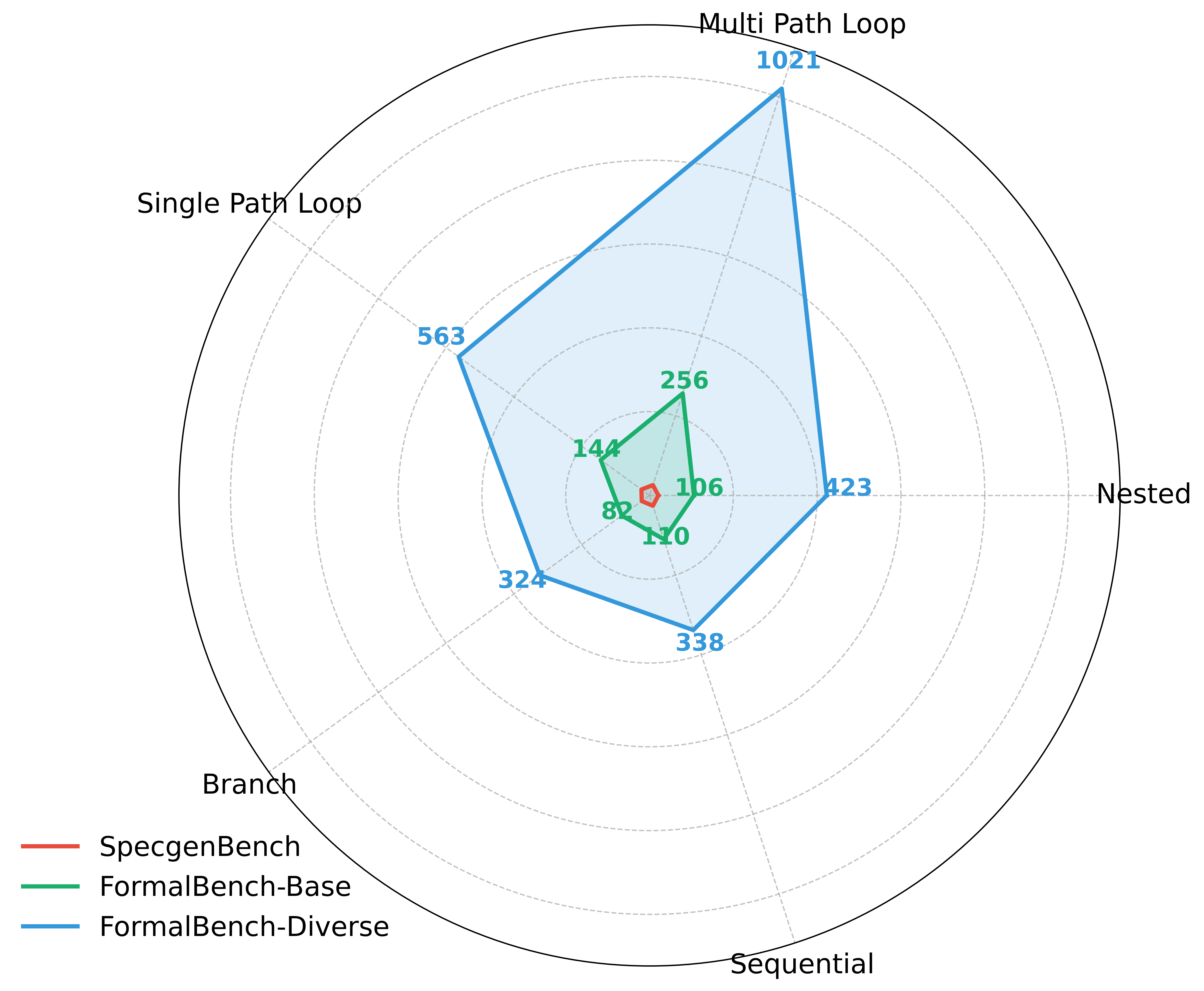}
  \caption{Distribution of our datasets and SpecGenBench over different control flow types.}
  \label{fig:data_category}
  \vspace{-5mm}
\end{figure}

The result is FormalBench-Base, a rigorously validated dataset of 700 programs with provably correct implementations. To ensure diversity, FormalBench-Base covers a wide range of control flow types, including sequential, branching, single-path loops, multi-path loops, and nested loop structures, as illustrated in Figure~\ref{fig:data_category}.

Finally, we apply 18 semantic preservation transformations from the literature~\cite{lecong2024reliableevaluationneuralprogram, rabin2021generalizability, zhang2023challenging} to FormalBench-Base, generating FormalBench-Diverse, a dataset of 6,219 program variants designed to evaluate the robustness of LLMs against syntactic variations. These transformations, detailed in Appendix~\ref{appx:transformations}, span multiple levels of code structure, including naming (e.g., variable renaming), expression (e.g., switching equal expressions) and statement (e.g., transforming switch statements to if-statements). 

\subsection{Evaluation Metrics}~\label{sec:metrics}

\textbf{Consistency Metrics.} As mentioned in Section~\ref{sec:statement}, we utilize deductive verification tools to determine the consistency of LLM-generated specifications. Unfortunately, in many cases, these verifiers return ``unknown'' either (1) when the specification lacks sufficient details to enable the verification tools to provide a final conclusion or (2) when the program or specification is overly complex so that these verifiers cannot provide a conclusion within time limits. In our experiments, we observe that most ``unknown'' results belong to the former situation, as programs in our benchmark have a manageable size so that the latter situation rarely happens. While prior works~\cite{wen2024enchanting, ma2024specgen} often merge \textit{unknown} cases with failures, we observe that specifications with this kind of outcome are qualitatively distinct from those that cause failures. Consequently, we classify ``unknown'' as a distinct outcome category in our study.

Therefore, our verification process produces three outcomes: 
(1) \textit{Verification Success}, where the implementation satisfies the specification; 
(2) \textit{Verification Failure}, where the implementation violates the specification; and 
(3) \textit{Unknown}, where the tool cannot definitively determine the result (e.g. due to timeouts or undecidability). 
Based on these categories, we define two consistency metrics as follows: (1) \textit{Success Rate (SR)}, the proportion of specifications that pass verification; and (2) \textit{Failure Rate (FR)}, the proportion of specifications that fail verification.

\textbf{Completeness Metrics.} Assessing the completeness of formal specifications is a well-recognized challenge. One potential approach involves evaluating their equivalence to manually written specifications. However, this method requires labor-intensive annotation of ground-truth specifications and the development of non-trivial techniques for quantifying semantic differences between specifications, making it both costly and complex.

In this work, we propose to use mutation testing as a proxy for measuring completeness of LLM-generated specifications. 
Particularly, we first apply mutation testing~\cite{andrews2005mutation} to generate a set of mutants, i.e., non-equivalent variants of the input programs created by deliberately injecting artificial faults. We then compute the fraction of these mutants that fail to satisfy the specification. This fraction is defined as the Completeness Rate (CR) of the specification. Mutation analysis is a well-established surrogate for evaluating the completeness of a test suite~\cite{andrews2005mutation, jia2010analysis}, functioning as a proxy for program specifications expressed through input-output examples. As such, mutation analysis offers a practical method for assessing the completeness of LLM-generated specifications.

To further validate its suitability, we compared our completeness metric with human judgment: annotators labeled 8 of 20 LLM-generated specifications as complete and 12 as incomplete. For the latter, they selected the more complete specification in 66 pairs. Our metric aligned with human evaluations, yielding higher scores for complete (0.91 $\pm$ 0.08) than for incomplete (0.77 $\pm$ 0.19) specifications and achieving a Cohen’s Kappa of 0.72, reflecting substantial agreement. These findings confirm that mutation analysis is a reliable measure of specification completeness.

\textbf{Robustness Metrics.} 
To evaluate the robustness of LLMs in synthesizing specifications under semantic-preserving transformations, we measure the \textit{Flip Rate (FlR)}, which captures cases where LLMs generate verifiable specifications for the original program $p$ but fail for its transformed versions. Moreover, we also measure the impact of unrobust behaviors on the performance of LLMs by measuring the success rates and failure rates on \textit{FormalBench-Diverse} (Section~\ref{sec:benchmark}). Since transformations may not apply universally, we normalize metrics over applicable transformations.

\textbf{Details.} We use OpenJML version 21.0 as our deductive verification tool and Major 3.0.1 as our mutation analysis tool. Full implementation details and formal formulations of these evaluation metrics are provided in Appendix~\ref{appx:metrics}. 

\section{Experiments}

In this section, we present our empirical results on LLMs using FormalBench, guided by the following research questions:

\begin{itemize}[noitemsep, topsep=0pt, left=0pt]
    \item \textbf{RQ}$_1$: \textit{How effective are LLMs in synthesizing formal specifications?}
    \item \textbf{RQ}$_2$: \textit{Can advanced prompting techniques improve the effectiveness of LLM?}
    \item \textbf{RQ}$_3$: \textit{How robust are LLMs in synthesizing formal specifications?}
    \item \textbf{RQ}$_4$: \textit{What are the common mistakes made by LLMs, and can they self-repair these errors?} 
\end{itemize}

Following prior works~\cite{ma2024specgen, flanagan2001houdini}, we focus on Java and its specification language, JML~\cite{leavens2006preliminary}, using OpenJML~\cite{cok2011openjml} as the verifier. LLMs are evaluated using their official chat templates and the same query prompts, as detailed in Appendix~\ref{appx:prompts}. To ensure fairness, we use sampling with a temperature setting of 0.7 across all LLMs. For open-source LLMs, the maximum number of tokens generated is limited to 2048 due to GPU constraints. Note that, as programs in our benchmark are relatively small, this context size does not reach the limit and requires truncations. A full description of the experimental setup is provided in Appendix~\ref{appx:exp_settings}.

\textbf{Cost Analysis.} Our experiments for closed-source LLMs cost approximately 250 USD, while those for open-source LLMs required around 100 GPU hours.

\begin{table*}[t]
  \centering
  \resizebox{0.75\textwidth}{!}{
  \begin{tabular}{llccc}
    \hline
    \textbf{}         & \textbf{Models}                 & \textbf{Success Rate (\%)} & \textbf{Failure Rate (\%)} & \textbf{Completeness (\%)} \\
    \midrule
    \multirow{22}{*}{\rotatebox[origin=c]{90}{\textbf{Open-Source LLMs}}} 
    & CodeQwen-1.5-7B         & 1.1   & 97.4  & 79.1  \\
    & + Few-shot prompt          & 3.9   & 85.6  & 74.1  \\ 
    \cline{2-5}
    & DeepSeek-R1-Qwen32B         & 3.0   & 94.1  & 98.2  \\
    & + Few-shot prompt          & 5.9   & 86.8  & 92.6  \\ 
    \cline{2-5}
    & Qwen2.5-32B         & 0.8   & 96.6  & 99.5  \\
    & + Few-shot prompt          & 6.2   & 84.3  & 91.9  \\ 
    \cline{2-5}
    & CodeQwen-2.5-32B          & 7.6   & 77.1  & 83.3  \\
    & + Few-shot prompt           & 11.4  & 66.8  & 82.1  \\
    & + COT                      & 9.2   & 69.4  & 86.8  \\
    & + LTM                      & 12.0  & 68.2  & 89.1  \\
    \cline{2-5}
    & DeepSeekCoder-33B-Instruct         & 2.7   & 88.8  & 77.0  \\
    & + Few-shot prompt            & 6.9   & 78.4  & 78.4  \\
    & + COT                      & 7.9   & 76.4  & 77.1  \\
    & + LTM                      & 8.9   & 70.4  & 81.6  \\
    \cline{2-5}
    & CodeLLaM-34B            & 0.1   & 99.6  & 100.0 \\
    & + Few-shot prompt             & 5.3   & 82.7  & 60.3  \\
    \cline{2-5}
    & LLama3-70B         & 5.7   & 83.4  & 88.9 \\
    & + Few-shot prompt          & 10.1 & 67.9  & 84.1  \\
    \cline{2-5}
    & DeepSeekR1-LLama70B         & 0.4  & 97.7  & 99.46 \\
    & + Few-shot prompt          & 4.0   & 82.7  & 98.7  \\
    \cline{2-5}
     & Qwen2.5-72B         & 3.0   & 87.8  & 94.1  \\
    & + Few-shot prompt          & 5.5   & 83.7  & 93.9  \\ 
    \midrule
    \multirow{18}{*}{\rotatebox[origin=c]{90}{\textbf{Proprietary LLMs}}} 
    & DeepSeek-V3-671B           & 8.4   & 65.2  & 89.6  \\
    & + Few-shot prompt            & 15.5  & 56.2  & 85.2  \\
    & + COT                      & 16.2  & 55.9  & 85.3  \\
    & + LTM                      & 16.6  & 56.8  & 89.6  \\
    \cline{2-5}
    & GPT-3.5               & 6.9   & 75.8  & 62.3  \\
    & + Few-shot prompt              & 12.6  & 59.8  & 59.0  \\
    \cline{2-5}
    & o3-mini               & 10.0   & 66.2  & 83.5  \\
    & + Low Reasoning             & 8.6  & 70.7  & 94.1  \\
    & + High Reasoning              & 10.6  & 65.6  & 97.2  \\
    & + Few-shot prompt              & 11.7  & 59.7  & 88.7  \\
    \cline{2-5}
    & GPT-4o               & 11.2  & 56.4  & 80.4  \\
    & + Few-shot prompt                & 13.4  & 56.4  & 77.6  \\
    & + COT                      & 13.4  & 61.5  & 81.2  \\
    & + LTM                      & 15.0  & 57.7  & 86.4  \\
    \cline{2-5}
    & Claude-3.5-Sonnet     & 10.5  & 64.5  & 91.2  \\
    & + Few-shot prompt   & 14.7  & 53.1  & 82.6  \\
    & + COT                      & 14.2  & 59.1  & 83.6  \\
    & + LTM                      & 15.4  & 51.1  & 86.4  \\
    \bottomrule
  \end{tabular}
  }
  \caption{\label{tab:rq1}
    Performance comparison of Open-Source and Commercial LLMs under zero-shot, in-context learning with few-shot prompt, chain-of-thought (COT), and least-to-most (LTM) prompting settings.
  }
\end{table*}

\subsection{\textbf{RQ}$_1$: Effectiveness of LLMs}

To answer the RQ$_1$, we evaluate the effectiveness of LLMs in synthesizing specifications by measuring their success rates, failure rates, and completeness rates. We assess eight popular open-source and proprietary LLMs on FormalBench-Base, as detailed in Appendix~\ref{appx:llms}. 
Detailed experimental results are presented in Table ~\ref{tab:rq1}.

\textbf{LLMs with zero-shot prompts.} From the results in Table~\ref{tab:rq1}, we observe that LLMs with zero-shot prompts perform poorly in synthesizing formal specifications, achieving a success rate of around 10\% and failure rates ranging from 56.4\% to 99.6\%. Most open-source LLMs, except for CodeQwen-2.5, exhibit particularly poor performance, with success rates below 3\%. Upon closer analysis, we found that the poor performance of open-source LLMs is caused by a lack of familiarity with JML syntax, resulting in a substantial number of invalid responses, up to 77\%. For example, these models often generate natural language descriptions instead of formal JML specifications, highlighting their inability to produce the formal grammar of JML without explicit guidance. 

\textbf{LLMs with few-shot prompts.}  We attempted to enhance the ability of LLMs to generate formal specifications by incorporating additional instructions on JML syntax and providing two pre-defined demonstration examples in the few-shot prompts, as shown in the Appendix~\ref{appx:prompts}. From the results in Table~\ref{tab:rq1}, we can see that this approach substantially improves LLM performance, with increases of up to 7.1 percentage points in success rates and reductions of up to 16.9 percentage points in failure rates. Furthermore, we observe a slight decline in completeness, although the overall completeness of the generated specifications remains high. This suggests a reasonable trade-off between correctness and consistency.
However, despite these improvements, the success rates remain relatively low at less than 16\%, highlighting inherent challenges of specification inference for LLMs.

\textbf{Reasoning LLMs.} We also found that reasoning LLMs, such as the o3-mini and R1 models, do not exhibit a higher success rate compared to other models. However, when their generated specifications are verifiable, they tend to demonstrate greater completeness, achieving approximately 90\%. Notably, for the o3-mini model, activating the high reasoning mode substantially increases completeness to 97.2\%, though this enhancement does not correspond to an improvement in the success rate. These results suggest that reasoning-focused training and fine-tuning yield benefits for LLMs in reasoning about program semantics.

\textbf{Open-source vs. Proprietary LLMs.} Moreover, we observe that proprietary LLMs such as DeepSeek-V3 and GPT-4o are substantially more effective than open-source LLMs. However, the best-performing open-source LLM, CodeQwen-2.5, shows considerable promise with a success rate of 12.0\%, only 3 percentage points lower than GPT-4o. This performance is particularly impressive given CodeQwen-2.5's compact size of 32B parameters. These findings suggest that open-source LLMs still have substantial potential for further advancement in this domain.

\textbf{Distribution over different Control-flow types.} Additionally, we analyze verification success and failure distributions across different control flow types (see detailed visualizations in Appendix~\ref{appx:distribution}).
From this analysis, we observe that LLMs are primarily capable of generating verifiable specifications for programs with simple control-flow structures, e.g., sequential or branched programs. 
However, they often struggle to synthesize formal specifications for programs that contain loops, where the complexity of control flow increases substantially, with a success rate of less than 10\% and failure rates of more than 50\%.
These findings highlight the limitations of LLMs in reasoning about complex control-flow structures that require advanced logical and inductive reasoning capabilities.

\begin{figure}
    \centering
    \includegraphics[width=\linewidth]{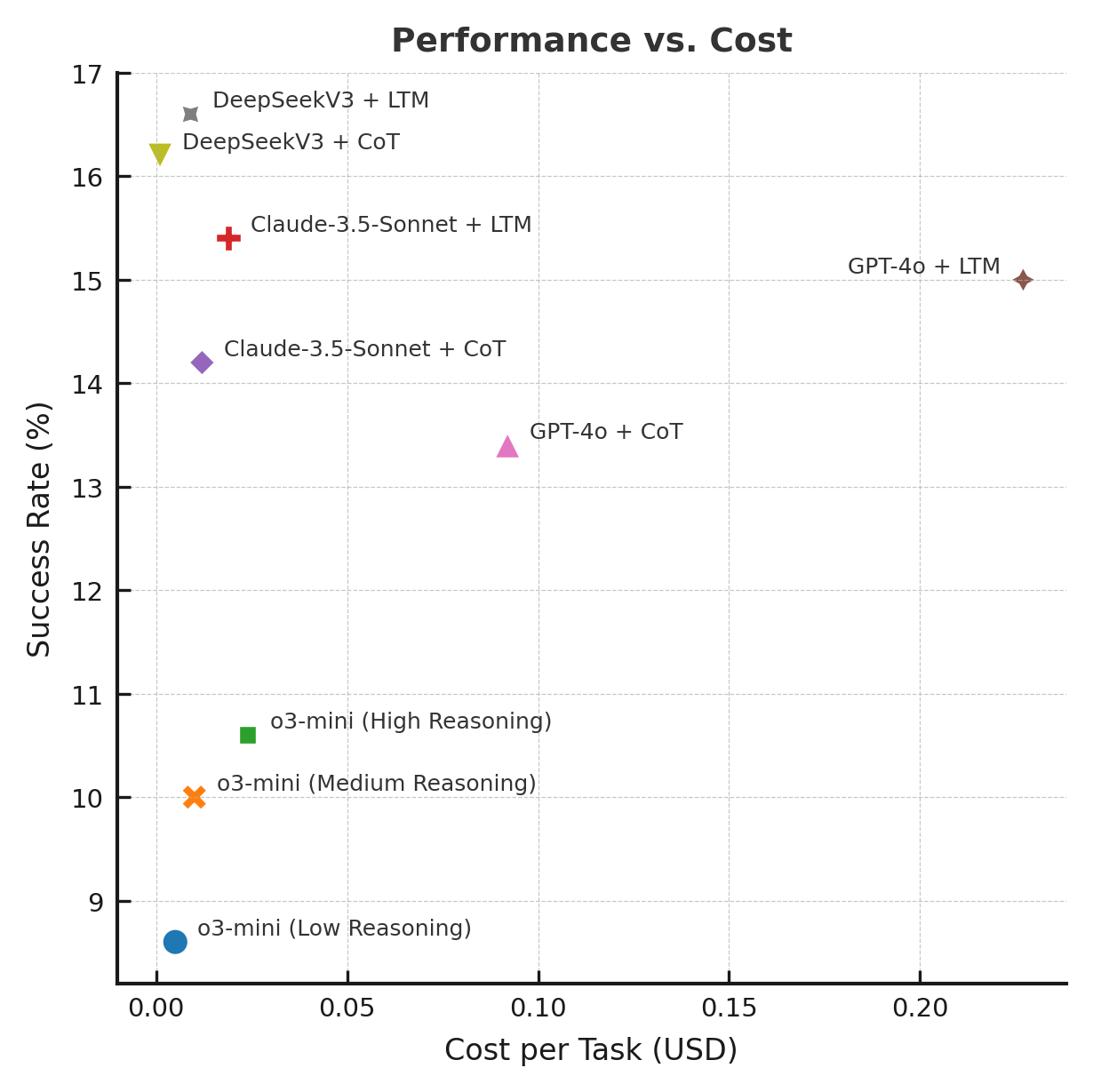}
    \caption{Performance versus cost per task of proprietary large language models (LLMs).}
    \label{fig:llm-performance-cost}
\end{figure}

\textbf{Performance versus Cost of Proprietary LLMs.} Finally, we analyzed the performance of proprietary LLMs relative to their cost per task.\footnote{Cost tracking was implemented retrospectively after several experiments, limiting the analysis to models with available statistics. Estimates are based on API pricing and token usage rather than direct provider data.} As shown in Figure~\ref{fig:llm-performance-cost}, DeepSeek-V3 models exhibit strong performance at low cost, indicating high cost-efficiency. GPT-4o performs well but incurs substantially higher costs, especially with the Least-to-Most prompt. In contrast, o3-mini models are more affordable among OpenAI models 
but achieve lower success rates. Moreover, while medium reasoning notably improves o3-mini’s performance over the low setting, the high reasoning mode offers minimal gains at nearly double the cost, raising concerns about its cost-effectiveness.

\subsection{\textbf{RQ}$_2$: Impact of Advanced Prompting}

To answer the RQ$_2$, we evaluate the top LLMs from RQ$_1$ (CodeQwen-2.5, DeepSeek-V2, DeepSeek-V3, GPT-4, and Claude 3.5 Sonnet) using two prompting techniques: chain-of-thought (CoT)\cite{kojima2022large} and least-to-most (LTM)\cite{zhou2022least}. Detailed prompt designs are in Appendix~\ref{appx:prompts}.

\textbf{Least-to-Most Prompts.} As shown in Table~\ref{tab:rq1}, LTM consistently improves the effectiveness of these LLMs, enhancing both consistency and completeness metrics. For example, the success rate of DeepSeek-V3 increases from 15.5\% with few-shot prompts to 16. 6\% with LTM (a 7\% improvement), while the completeness rate increases from 85. 2\% to 89. 6\% (a 5\% improvement). These improvements are observed not only in proprietary LLMs, but also in open-source LLMs. Specifically, the success rates of CodeQwen-2.5 and DeepSeek-V2 improve substantially, from 11.4\% and 6.9\% to 12.0\% and 8.9\%, respectively. Overall, these findings suggest that LTM prompting, when combined with few-shot demonstrations, should be used to optimize the effectiveness of LLMs in synthesizing program specifications.

\textbf{Chain-of-Thought Prompts.} In contrast, the impact of CoT on LLMs is mixed, with both positive and negative outcomes. For example, CoT improves the success rate of DeepSeek-V3 from 15.5\% to 16.2\%. However, it has no effect on the success rate of GPT-4o and even decreases the performance of the Claude 3.5 Sonnet. CoT even substantially increases the failure rates of GPT-4o and Claude-3.5-Sonnet from 56.4\% and 53.1\% to 61.5\% and 59.1\%.
We suspect that this is because CoT relies on the model's ability to self-reason, while LTM provides human-instructed reasoning steps and explicit demonstrations, which guide the models toward better reasoning.

\begin{table*}[t]
  \centering
  \resizebox{0.75\textwidth}{!}{
  \begin{tabular}{lccccc}
    \hline
    \textbf{Model Name} & \multicolumn{2}{c}{\textbf{FormalBench-Base}} & \multicolumn{2}{c}{\textbf{FormalBench-Diverse-N}} & \textbf{Flip Rate (\%)} \\
    \cline{2-5}
    & \textbf{SR (\%)} & \textbf{FR (\%)} & \textbf{SR (\%)} & \textbf{FR (\%)} & \\
    \midrule
    DeepSeek-V3 & 9.3 & 62.1 & 7.8 & 65.0 & 27.2 \\
    Claude-3.5-Sonnet & 10.0 & 64.1 & 8.3 & 64.3 & 39.20 \\
    GPT-4o & 11.9 & 60.1 & 10.9 & 64.1 & 29.2 \\
    \bottomrule
  \end{tabular}
  }
  \caption{\label{tab:rq3}
    Robustness evaluation of LLMs on FormalBench-Base and FormalBench-Diverse-N benchmarks using different metrics Success Rate (SR), Failure Rate (FR), and Flip Rate.
  }
\end{table*}

\textbf{Effectiveness on Complex Control-Flow Programs.} While LTM prompting can further improve the performance of LLMs, these improvements are primarily observed in reasoning programs involving branching and sequential control flow. In contrast, the impact of LTM prompting on improving reasoning for programs with complex control flow, such as those containing loops, remains unclear. As a result, the performance of LLMs in loop-containing programs remains low, with success rates of less than 10\%. This further underscores the limitations of LLMs in reasoning about programs with loops, which require more advanced inductive reasoning capabilities.

\subsection{\textbf{RQ}$_3$: Robustness of LLMs}

To answer the RQ$_{3}$, we assess LLM robustness by evaluating their performance on semantically equivalent but syntactically diverse programs using FormalBench-Diverse-N, a subset of 1,794 \textit{natural} program transformations~\cite{lecong2024reliableevaluationneuralprogram} from FormalBench-Diverse (see Appendix~\ref{appx:diversenatural}). Due to resource constraints, we focus on the top three LLMs: GPT-4, Claude 3.5, and DeepSeek-V3.

Our experimental results, presented in Table~\ref{tab:rq3}, reveal substantial robustness challenges for all evaluated LLMs. Specifically, we observe flip rates, the proportion of semantically equivalent programs for which the model does not generate verifiable specifications, ranging from 27. 2\% to 39. 2\%. Among the models, Claude-3.5-Sonnet is the most severely impacted, with a flip rate of 39.2\%, indicating that it does not generate verifiable specifications for nearly 40\% of the transformations when the original programs had verifiable specifications.

More critically, this lack of robustness leads to a notable decrease in success rates and an increase in failure rates. For instance, the success rate of DeepSeek-V3 drops from 9.3\% to 7.8\%, a 16\% reduction, while GPT-4o and Claude-3.5-Sonnet experience reductions of 9.5\% and 17\%, respectively. Similarly, failure rates increase by up to 6.6\%, further underscoring the sensitivity of LLMs to the syntactic variation created by semantic-preserving transformations.

\begin{figure*}[t]
    \centering
    \begin{subfigure}[b]{0.32\textwidth}
        \centering
        \includegraphics[width=\textwidth]{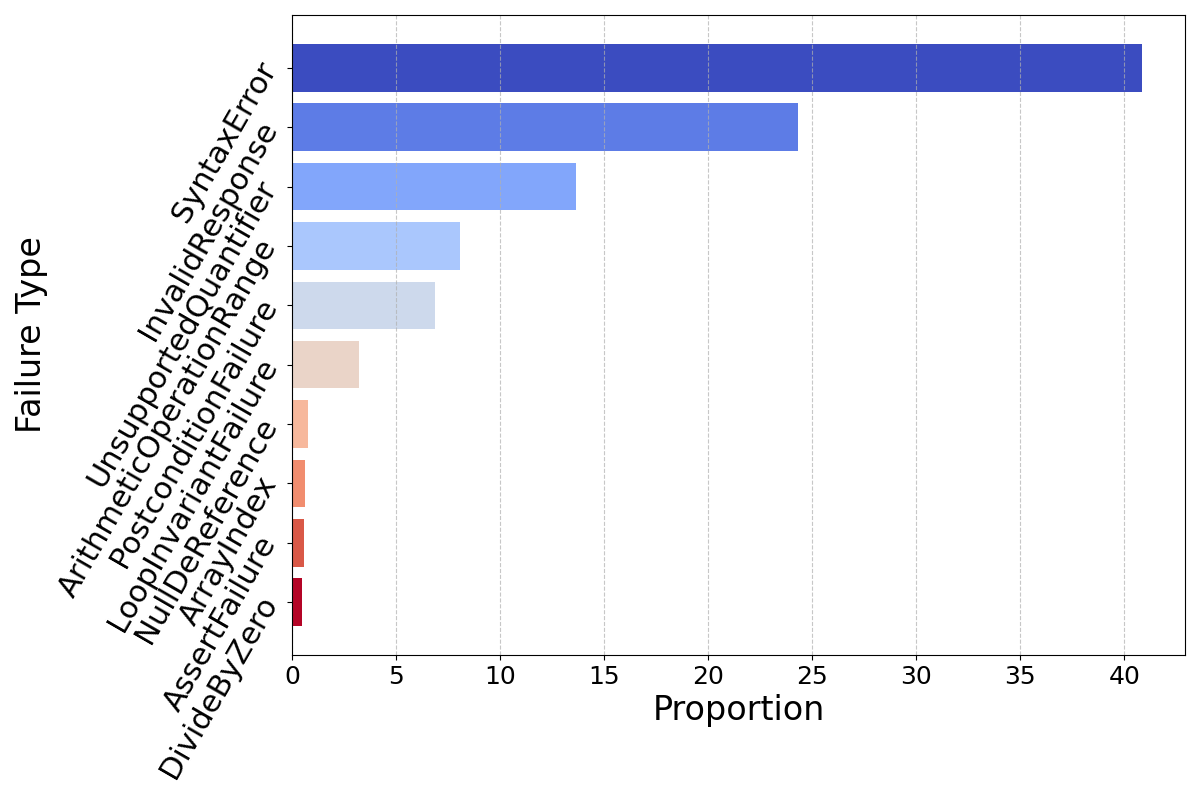}
        \caption{LLMs with Zero-shot prompts}
    \end{subfigure}
    \hfill
    \begin{subfigure}[b]{0.32\textwidth}
        \centering
        \includegraphics[width=\textwidth]{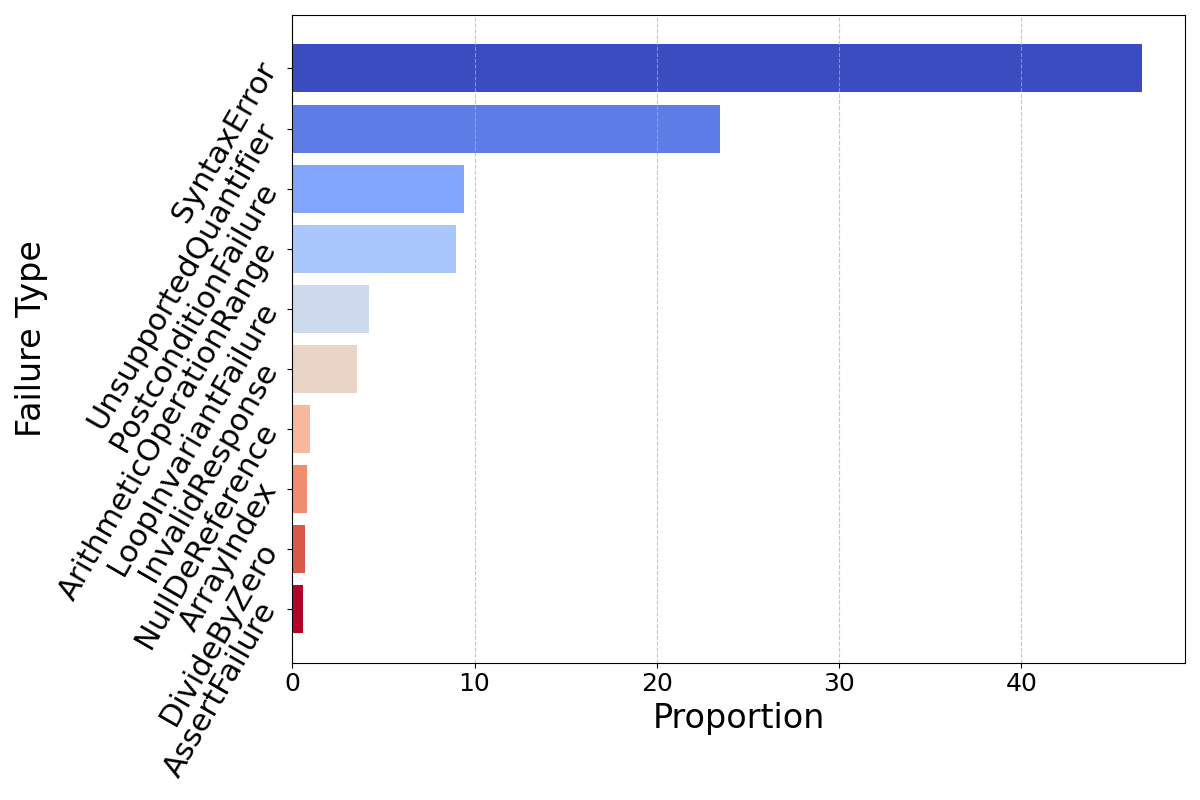}
        \caption{LLMs with Few-shot prompts}
    \end{subfigure}
    \hfill
    \begin{subfigure}[b]{0.32\textwidth}
        \centering
        \includegraphics[width=\textwidth]{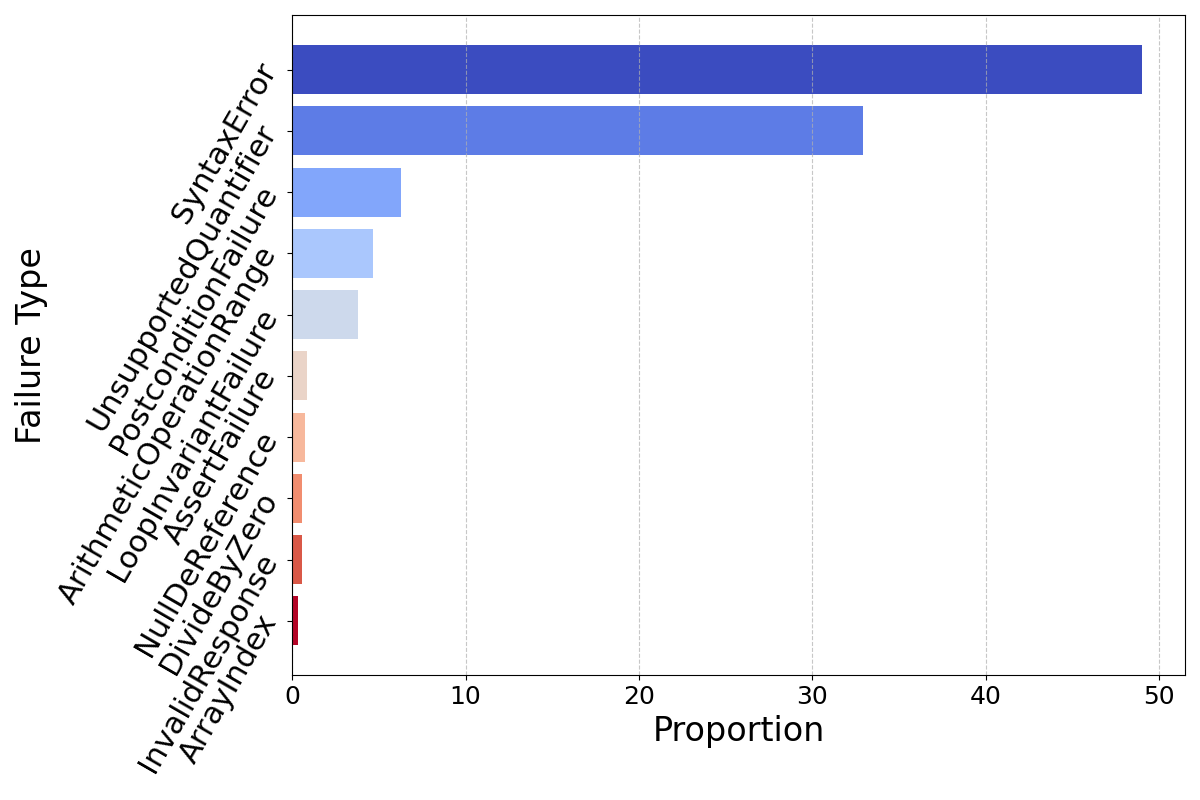}
        \caption{LLMs with LTM prompts}
    \end{subfigure}
    \caption{Top-10 failure category of LLMs with various prompts}
    \label{fig:failure_category}
\end{figure*}

These findings highlight the limited robustness of LLMs against semantic-preserving transformations, which expose a critical dependence on syntactic patterns rather than underlying semantic properties. This indicates that current LLMs still lack the deep semantic reasoning capabilities necessary to generalize across functionally equivalent but syntactically varied programs.

\subsection{\textbf{RQ}$_4$: Common Failures and Self-Repair Ability of LLMs}

To answer the RQ$_4$, we begin by conducting a semi-automated analysis, as outlined in Appendix~\ref{appx:failure_analysis}, to categorize the failures of LLMs. For each type of failure, we sample a subset of instances and investigate their root causes. Building on these insights, we design customized prompts that include failure descriptions, additional guidance, and illustrative examples to enable LLMs to self-repair these errors. Additional details on repair prompts are provided in the Appendix~\ref{appx:repair_prompts}.

\subsubsection{Common Failures}

In total, we identified 32 failures of LLMs based on their error messages.  Figure~\ref{fig:failure_category} illustrates the 10 most common failure categories encountered across llm when using zero-shot, few-shot, and least-to-most (LTM) prompts. 

\begin{figure*}[t]
    \centering
    \begin{subfigure}[b]{0.32\textwidth}
        \centering
        \includegraphics[width=\textwidth]{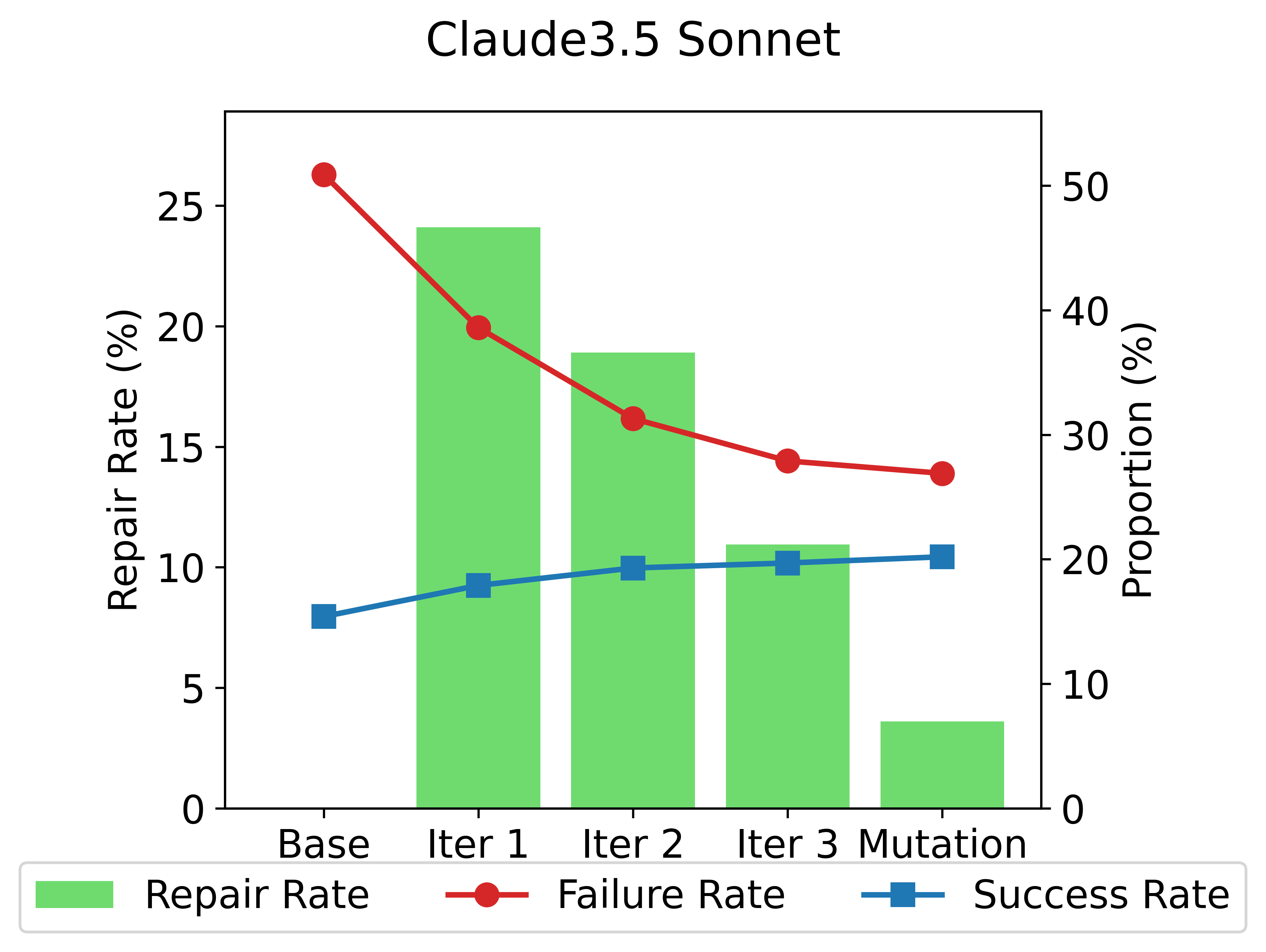}
        \caption{Claude-3.5-Sonnet}
    \end{subfigure}
    \hfill
    \begin{subfigure}[b]{0.32\textwidth}
        \centering
        \includegraphics[width=\textwidth]{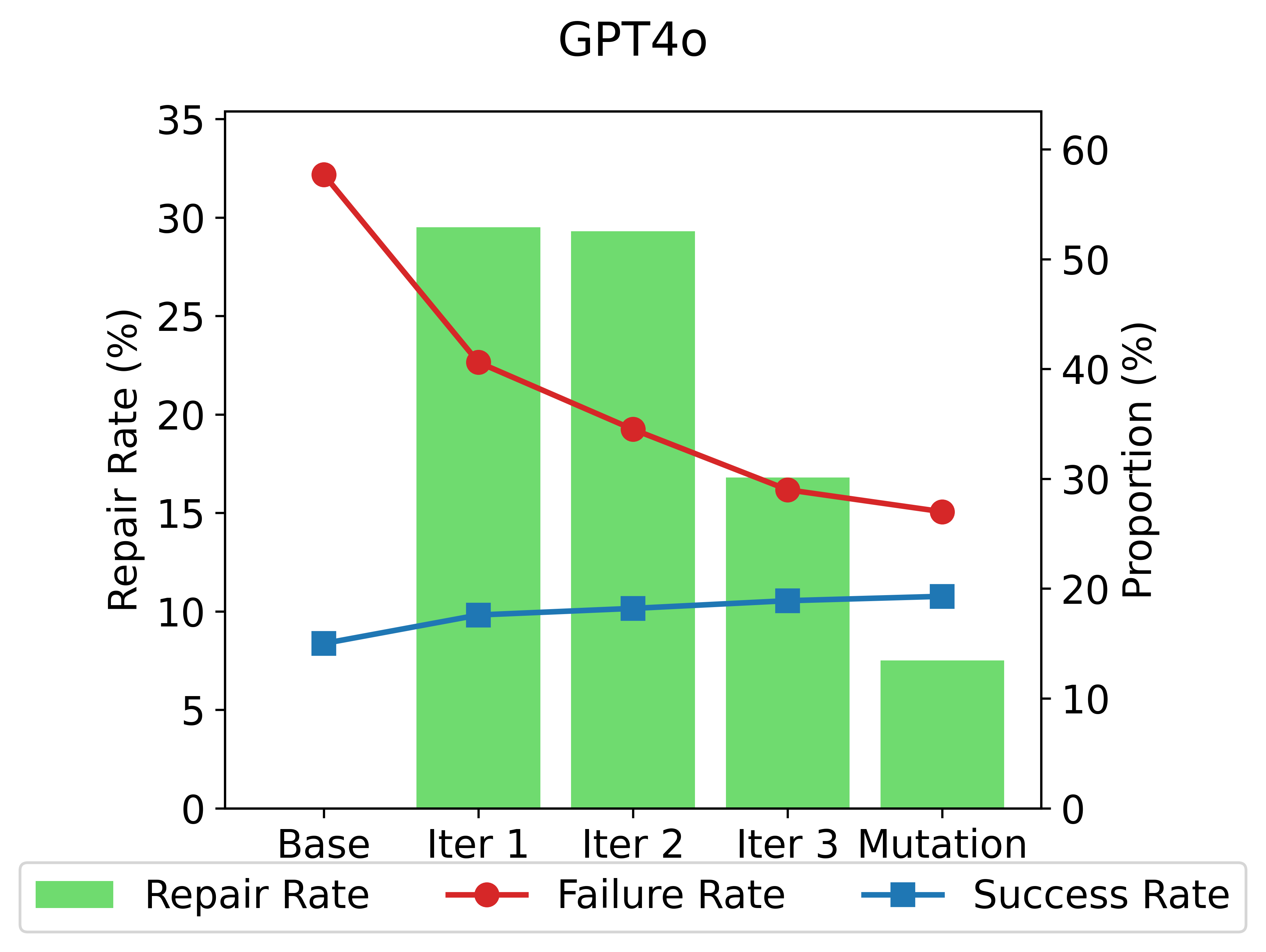}
        \caption{GPT-4o}
    \end{subfigure}
    \hfill
    \begin{subfigure}[b]{0.32\textwidth}
        \centering
        \includegraphics[width=\textwidth]{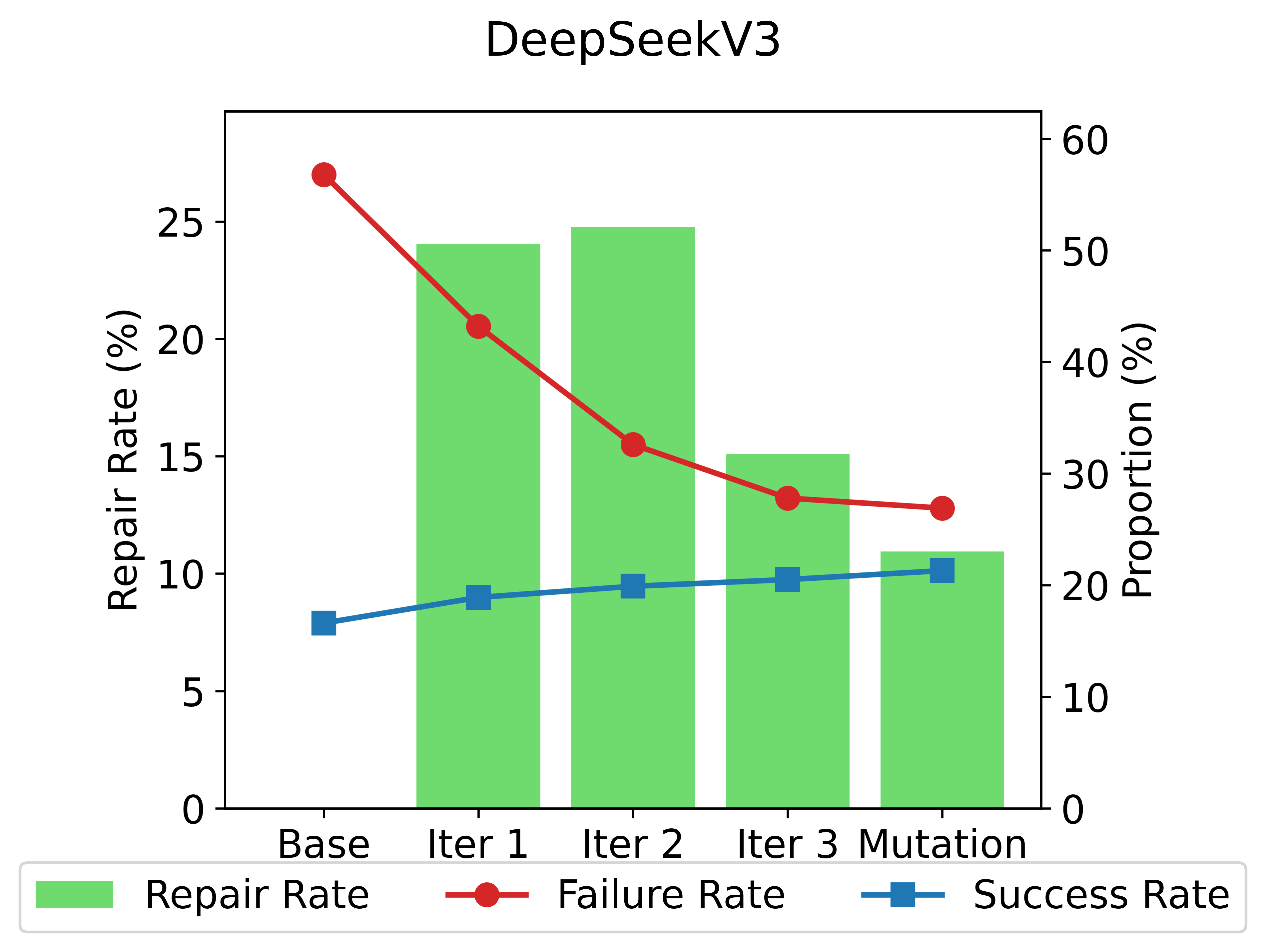}
        \caption{DeepSeekV3}
    \end{subfigure}
    \caption{Effectiveness and self-repair rates of LLMs across iterations: ``Iter $i$'' represents self-repair with feedback, while   ``mutation'' represents results from mutation-based repair in the final iteration.}
    \label{fig:repair}
\end{figure*}

\textbf{Syntax Errors.} Among failure types, ``SyntaxError'' is the most frequent, and LLMs often generate specifications that violate the JML or Java syntax. This issue persists in most models. A common example is the error message “Unexpected or misspelled JML token,” which occurs when an LLM produces incorrect JML grammar. This highlights the challenge of expressing implicit program intent in formal languages, a key distinction from natural language specifications such as code comments. More critically, LLMs with zero-shot prompts yield about 25\% invalid responses (e.g., producing Javadoc comments instead of JML). Fortunately, this rate drops to 5\% with few-shot prompts and further to 1\% with LTM prompts.

\textbf{Reasoning Errors.} LLMs often encounter reasoning errors, particularly with quantifiers, postconditions, loop invariants, and arithmetic bounds. The most frequent are ``UnsupportedQuantifier'' errors.
In these cases, LLMs rely on inductive quantifiers such as \textbackslash sum or \textbackslash product, which are not supported by deductive verification for reasoning about program behaviors. In practice, formal experts must supplement these quantifiers with auxiliary mathematical functions and lemmas to enable inductive reasoning. For improved formal specification synthesis, LLMs must adopt similar human-like strategies rather than relying on unsupported quantifiers.

Following ``UnsupportedQuantifier'', the next most common failure categories, ``PostconditionFailure'', ``LoopInvariantFailure'', and ``ArithmeticOperationRange'', account for nearly 30\% of total failures. These errors occur when the verification tool cannot prove postconditions, loop invariants, or arithmetic bounds (e.g., to prevent overflow). Our analysis identifies three main root causes: (1) incorrect specifications, (2) weak or incorrect preconditions that render specifications unprovable, and (3) incomplete reasoning about program behavior, leaving verifiers with insufficient information. These findings underscore the need for LLMs to enhance their reasoning capabilities for more effective formal specification synthesis.

\subsubsection{Self-Repair}

To evaluate LLM's self-repair ability, we (1) develop a simple failure classifier using pattern matching and (2) design customized prompts with failure descriptions, guidance, and examples (see Appendix~\ref{appx:prompts}). Additionally, we assess SpecGen’s mutation-based repair~\cite{ma2024specgen} for verification failures. Due to resource constraints, we evaluate only the top three LLMs: Claude-3.5-Sonnet, GPT-4o, and DeepSeek-V3, and present the results in Figure~\ref{fig:repair}.

The results show that LLMs effectively repair errors using our custom prompts, improving success rates by 25\%, from 16\% to 20\%, and reducing failure rates from over 50\% to under 30\%. Mutation-based repair further increases success by 0.5 percentage points and reduces failure by 1 to 2 percentage points. Additionally, LLMs can also self-repair across various error categories, such as fixing 53.7\% of ``SyntaxErrors'', 79\% of ``LoopInvariantFailures'', and 65\% of ``PostconditionFailures'' in the first iterations. This pattern persists across subsequent iterations, highlighting the flexibility of self-repair. In contrast, mutation-based repair is limited to specific errors, such as not addressing ``ArithmeticOperationRange'' errors. Notably, both methods preserve the completeness of generated specifications, improving the number of verifiable specifications without sacrificing quality. However, we identify two limitations of self-repair approaches. First, self-repair rates decrease with each iteration, leading to saturation in success and failure rates. Second, mutation-based repair is computationally expensive and requires frequent calls to verification tools, so it should be used sparingly, ideally as a final step, to minimize costs.

\section{Related Works}

\textbf{Reasoning Evaluation of LLMs.} Numerous datasets have been curated to evaluate the reasoning capabilities of LLMs across diverse domains, including mathematical~\cite{cobbe2021training,hendrycks2021measuring}, logical~\cite{liu2021logiqa, yang2022language}, and causal reasoning~\cite{jin2024cladder, jin2023can}. Recent work explores code reasoning, evaluating LLMs' ability on program semantic inference.
~\cite{hu2018deep, jain2024livecodebench, chen2024reasoning}. Early studies focus on code summarization~\cite{husain2019codesearchnet, hu2018deep}, capturing high-level understanding rather than fine-grained semantic reasoning. Recent studies examine code reasoning in detail through output prediction~\cite{jain2024livecodebench}, execution trace simulation~\cite{chen2024reasoning}, and invariant inference~\cite{pei2023can}, yet they still address only partial program semantics. In contrast, FormalBench targets formal specification inference, demanding exhaustive reasoning that produces precise, verifiable specifications for every possible execution.

\textbf{Formal Specification Inference.}  Traditional dynamic analysis methods, such as Daikon~\cite{ernst2007daikon}, Houdini~\cite{flanagan2001houdini}, and DIG~\cite{nguyen2014dig}, infer likely invariants from observed behaviors using predefined templates. However, these tools often yield trivial invariants (e.g., nums != null) and struggle with complex functional properties~\cite{ma2024specgen}. Recent work leverages LLMs to address these limitations. Early approaches~\cite{pei2023can, chakraborty2023ranking} fine-tuned LLMs for invariant inference but focused on specific cases, such as loop invariants or unverified likely invariants. Nilizadeh et al.\cite{nilizadeh2021exploring} manually crafted complete program specifications to assess automated repair effectiveness. Building on this, newer methods such as SpecGen\cite{ma2024specgen} and AutoSpec~\cite{wen2024enchanting} automatically generate full formal specifications via iterative refinement and static analysis. However, as discussed in Section~\ref{sec:intro}, their evaluations remain limited, highlighting the need for FormalBench and more comprehensive assessments of LLM effectiveness.

\section{Conclusion}

In this work, we introduce FormalBench, a comprehensive and large-scale benchmark for specification inference. FormalBench integrates robust evaluation metrics to assess the consistency, completeness, and robustness of LLMs on this task. Using FormalBench, we conduct an extensive evaluation of 14 popular LLMs, revealing their limited effectiveness and lack of robustness in synthesizing formal specifications, even with advanced prompting techniques. We further analyze their common failure patterns and propose a set of customized prompts, leveraging LLMs' self-repair capabilities to enhance their performance. Overall, FormalBench aims to enable a thorough evaluation and deeper understanding of LLMs in formal program reasoning. 

\section{Limitations.}
The limitations of this work are as follows:

First, mutation analysis, which sits at the core of our completeness metric, relies on predefined rules to systematically break program behavior, assuming that the generated mutants will exhibit semantic differences from the original program, thereby triggering detectable errors. However, the presence of equivalent mutants, i.e., semantically identical variants of the original program, presents a challenge, as they evade detection, leading to false positives and undermining the accuracy of completeness metrics. To mitigate this issue, we incorporate Equivalent Mutant Suppression (EMS)~\cite{kushigian2024equivalent}, a state-of-the-art technique for filtering out equivalent mutants. To evaluate the effectiveness of EMS, we manually inspected 20 randomly selected programs from our benchmark. Of the 232 mutants, only 3 (1.3\%) were equivalent, indicating low noise. While EMS reduces the prevalence of equivalent mutants, some undetected equivalent mutants may still affect the validity of our results. Future research should focus on the development of metrics that more effectively measure the completeness of generated specifications, thereby reducing reliance on mutation analysis as an isolated proxy.

Second, our experiments produced a substantial number of "unknown" results from the program verification tools. Manual inspection suggests that these cases are generally associated with higher-quality specifications compared to those with verification failures; yet, they introduce ambiguity due to inherent limitations in deductive verification tools. Future work should prioritize techniques for interpreting or eliminating these ambiguous results, possibly through enhanced symbolic execution or dynamic verification methods.

Third, our study focuses only on Java and the Java Modeling Language due to their popularity. While it is possible to extend our approach to other programming languages such Python or C, we focus on Java and JML due to the maturity of its associated tooling. For instance, mutation analysis for Java is well-established in this ecosystem, enabling high-quality analysis of completeness. Additionally, Java supports JML annotations, allowing specifications to be embedded directly within the code, unlike languages such as Python, which lack dedicated specification languages. Compared to specification-aware languages like Dafny, JML enables the specification of a general-purpose programming language, i.e., Java, without requiring a shift to a verification-aware programming language. Overall, it offers the most practical and effective framework for our study. We acknowledge this limitation and will generalize our approach to other languages in future work.

Finally, our experiments did not include OpenAI-o1 and DeepSeekR1, the latest LLMs at the time of writing. For OpenAI-o1, the associated costs were prohibitively high, so we could not incorporate it into our experiments due to resource constraints. As an alternative, we conducted experiments on o3-mini, the latest reasoning model from OpenAI, with reasonable cost. 
For DeepSeekR1, we currently cannot access DeepSeek models due to a recently introduced policy within our university that prohibits their use. Instead, we deployed the best open-source DeepSeek-R1 models under resource constraints of the NVIDIA A100 80G: DeepSeek-R1-Distill-Qwen-32B and DeepSeek-R1-Distill-LLama-70B.

\bibliography{main}

\clearpage

\appendix

\section{Semantic-preserving Transformations}
\label{appx:transformations}

In this study, we curate a set of 18 semantic-preserving transformations from recent studies~\cite{lecong2024reliableevaluationneuralprogram, zhang2023challenging, rabin2021generalizability} including:

\begin{itemize}
    \item \textbf{VariableRenaming-1} replaces a variable name by its first characters;
    \item \textbf{VariableRenaming-1} replaces a variable name by substitutions derived from CodeBERT~\cite{feng2020codebert};
    \item \textbf{SwitchRelation} transforms relational expressions by swapping the operands. For example, the expression \texttt{a < b} is transformed into \texttt{b > a}.
    \item \textbf{Unary2Add} modifies unary operations or increments by converting them into normal assignment statements. For instance, \texttt{i++;} is transformed into \texttt{i = i + 1;}.
    \item \textbf{Add2Equal} converts add/subtract assignments into equal assignments. For example, \texttt{a += 9;} is transformed into \texttt{a = a + 9;}, and \texttt{b -= 10;} is transformed into \texttt{b = b - 10;}.
    \item \textbf{MergeVarDecl} merges multiple variable declarations into a single statement. For instance, \texttt{int a;} and \texttt{int b;} are merged into \texttt{int a, b;}.
    \item \textbf{InfixDividing} divides an in/pre/post-fix expression into two separate expressions, storing intermediate results in a temporary variable. For example, \texttt{x = a + b * c} is transformed into \texttt{temp = b * c;} followed by \texttt{x = a + temp}.
    \item \textbf{SwitchEqualExp} switches the two expressions on both sides of an infix expression where the operator is \texttt{=}. For instance, \texttt{a == b} is transformed into \texttt{b == a}.
    \item \textbf{SwitchStringEqual} switches the order of string equality checks. For example, \texttt{a.equals(b)} is transformed into \texttt{b.equals(a)}.
    \item \textbf{For2While} transforms a for-loop into a while-loop, restructuring the loop for different control flow requirements.
    \item \textbf{While2For} transforms a while-loop into a for-loop;
    \item \textbf{ElseIf2If} transforms an If...Else if... structure into a nested If...Else structure;
    \item \textbf{Switch2If} transforms a Switch-Case structure into an If-Else structure, converting switch-based logic into a series of conditional checks.
    \item \textbf{SwapStatement} swaps two statements that have no control or data dependency;
    \item \textbf{ReverseIf} switches the code blocks in the if statement and the corresponding else statement, inverting the condition and its associated logic.
    \item \textbf{If2CondExp} changes a single if statement into a conditional expression statement, simplifying the code into a more concise form. For example, \texttt{if (condition) \{ StatementA \} else \{ StatementB \}} becomes \texttt{condition ? StatementA : StatementB}.
    \item \textbf{CondExp2If} changes a conditional expression statement into a single if statement. For example, \texttt{condition ? StatementA : StatementB} becomes \texttt{if (condition) \{ StatementA \} else \{ StatementB \}}.
    \item \textbf{DividingComposedIf} divides an if statement with a compound condition \((\wedge, \vee, -)\) into two nested if-statements, breaking down complex conditions into simpler, more manageable parts.
\end{itemize}

\section{Evaluation Metrics}
\label{appx:metrics}

In this appendix, we present the formal definition and implementation details of our evaluation metrics, presented in Section~\ref{sec:metrics} 

\subsection{Consistency Metrics}
Given a benchmark dataset $\mathcal{D}$, an LLM $\mathcal{L}$, and a verification tool $\mathcal{V}$, success rate (SR) and failure rate (FR) are formally defined as follows:

\begin{equation*}
    SR(\mathcal{L}) = \dfrac{\big|\{r \in \mathcal{D} \mid g = \mathcal{L}(r) \wedge \mathcal{V}(g, r) = \text{ok}\}\big|}{|\mathcal{D}|},
\end{equation*}

\begin{equation*}
    FR(\mathcal{L}) = \dfrac{\big|\{r \in \mathcal{D} \mid g = \mathcal{L}(r) \wedge \mathcal{V}(g, r) = \text{fail}\}\big|}{|\mathcal{D}|},
\end{equation*}

where $g = \mathcal{L}(p)$ is the specification generated by $\mathcal{L}$ for the reference program $r$, and $\mathcal{V}(g, p)$ denotes the verification result of $g$ on $p$ using $\mathcal{V}$. To ensure the consistency between the generated specification and the reference program, we employ OpenJML~\cite{cok2011openjml}, a widely used program verification tool. Specifically, we utilize its latest version (21.0) in the \texttt{esc} mode (Extended Static Checker~\cite{flanagan2002extended}) with the CVC4 SMT solver~\cite{barrett2011cvc4}. Additionally, we enable arithmetic mode and assume that pointers are nullable by default.

\subsection{Completeness Metrics}
For a generated specification $g$ and a reference program $r$, the completeness rate (CR) is formally defined as follows:

\begin{equation*}
    CR(g, r) = \dfrac{\big|\{p \in \mathcal{P}(r) \mid \mathcal{V}(g, p) \neq \text{ok}\}\big|}{|\mathcal{P}(r)|},
\end{equation*}

where $\mathcal{P}(r)$ is the set of mutants for $r$, and $\mathcal{V}(g, p)$ is the verification result of $g$ on mutant $p$. Higher CR indicates greater completeness, as $g$ detects more faults. To generate s$\mathcal{P}(r)$, we utilize Major~\cite{just2014major}, a widely recognized mutation testing framework, using its latest version (3.0.1). To mitigate the generation of equivalent mutants, we further employ EMS~\cite{kushigian2024equivalent}, a state-of-the-art equivalent mutant suppression technique.

\subsection{Robustness Metrics}

To evaluate the robustness of LLMs, we leverage a set of 18 semantic-preserving transformations, presented in Section~\ref{appx:transformations}. Given $p$, its set of transformed programs $\mathcal{T}$, LLM $\mathcal{L}$, and verification tool $\mathcal{V}$, with $g = \mathcal{L}(p)$ verified as correct ($\mathcal{V}(g, p) = \text{ok}$), Flip Rate (FlR) is defined as follows:

\begin{equation*}
    FlR(p, \mathcal{T}) = \dfrac{\big|\{t \in \mathcal{T} \mid g' = \mathcal{L}(t) \wedge \mathcal{V}(g', t) \neq \text{ok}\}\big|}{|\mathcal{T}|}.
\end{equation*}

Moreover, we also measure the consistency and completeness metrics of LLMs on our transformed dataset. Since transformations may not apply universally, we normalize metrics over applicable transformations. For a reference program $r$, its set of transformed programs $\mathcal{T}$, and metric $\mathcal{M}$, the normalized metric $\mathcal{M'}$ is defined as follows:
\begin{equation*}
    \mathcal{M'}(\mathcal{T}) = \dfrac{\sum_{t \in \mathcal{T}} \mathcal{M}(t)}{|\mathcal{T}|},
\end{equation*}
where $\mathcal{M}$ can be $SR$, $FR$, or $CR$. 

\section{Evaluated Large Language Models}
\label{appx:llms}

We evaluate the following models with our FormalBench benchmark:
\begin{itemize}
    \item \textbf{Open-source LLMs:}
    \begin{itemize}
        \item CodeQwen-1.5: CodeQwen-1.5-7B~\cite{qwen}
        \item CodeQwen-2.5: Qwen2.5-Coder-32B-Instruct~\cite{qwen2, hui2024qwen2}
        \item CodeLLama: CodeLlama-34b-Instruct-hf~\cite{roziere2023code}
        \item DeepSeekCoder~\cite{deepseek-coder}
    \end{itemize}
    \item \textbf{Proprietary LLMs: }
    \begin{itemize}
        \item DeepSeek-V3-671B~\cite{liu2024deepseekv3}
        \item GPT3.5: GPT-3.5-turbo~\cite{openai_gpt35_turbo}
        \item GPT-4o~\cite{openai_gpt4o}
        \item Claude: Claude-3.5-Sonnet~\cite{claude}
    \end{itemize}
\end{itemize}

\section{Experimental Settings}
\label{appx:exp_settings}

To query LLMs, we implemented our framework using LangChain, an open-source framework designed to streamline the development of applications leveraging llm. For running open-source LLMs, we use an NVIDIA A100 GPU with 80GB of VRAM and an Intel® Xeon® Gold 6326 CPU operating at 2.90 GHz. For running the verification tool, we leverage an Intel® Xeon® Platinum 8358 CPU operating at 2.90 GHz with 28 CPU cores and 1953GB of RAM.

\section{Constructions of FormalBench-Diverse-Natural}
\label{appx:diversenatural}

To construct FormalBench-Diverse-Natural, we first assess the naturalness of semantic-preserving transformations by measuring the relative change in cross-entropy, following established methods in previous studies~\cite{lecong2024reliableevaluationneuralprogram,ray2016naturalness,hindle2016naturalness}. We then select 50\% of the transformations, sorted by naturalness score. Finally, we choose programs with at least three transformations to avoid bias in the calculation of normalized metrics.

\section{Distribution of Verification success and failures over different control-flow types.}
\label{appx:distribution}

In this section, we present the distribution of verification success and failures over different control-flow types for three evaluated LLMs.

\begin{figure}[h]
    \centering
    \includegraphics[width=0.85\linewidth]{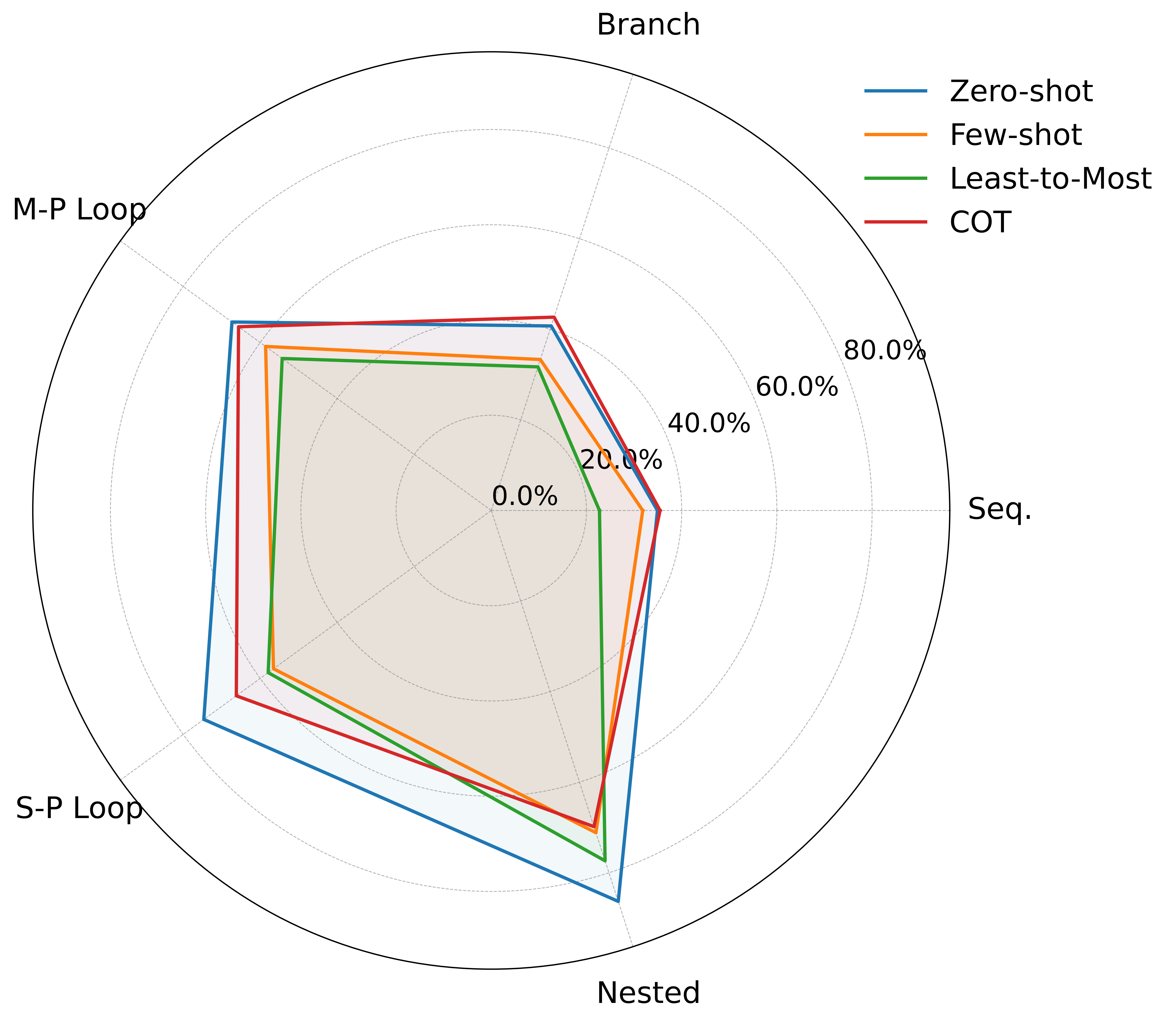}
    \caption{Verification Failures of Claude-3.5-Sonnet}
    \label{fig:enter-label}
\end{figure}

\begin{figure}[h]
    \centering
    \includegraphics[width=0.85\linewidth]{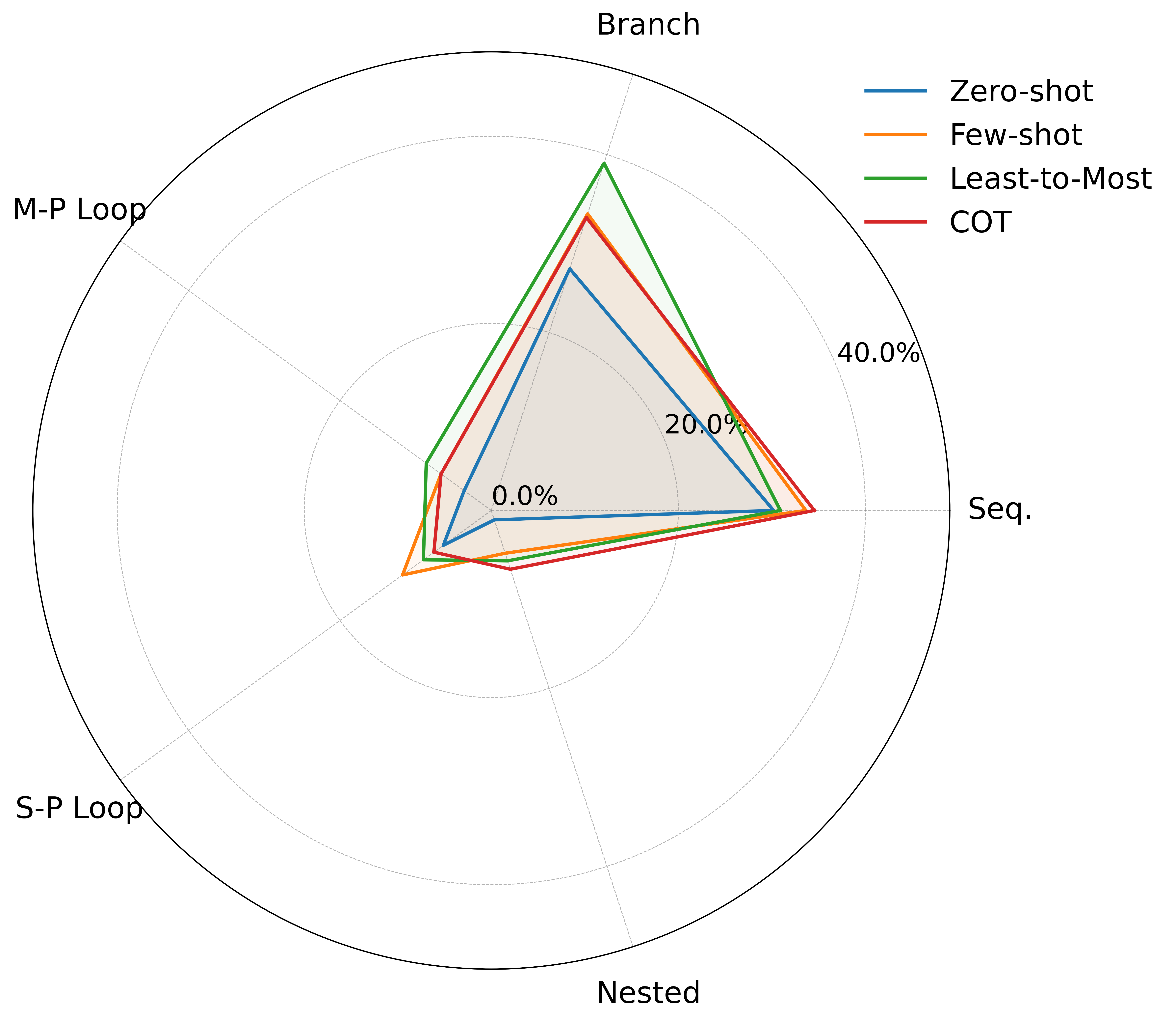}
    \caption{Verification Successes of Claude-3.5-Sonnet}
    \label{fig:enter-label}
\end{figure}

\begin{figure}[h]
    \centering
    \includegraphics[width=0.85\linewidth]{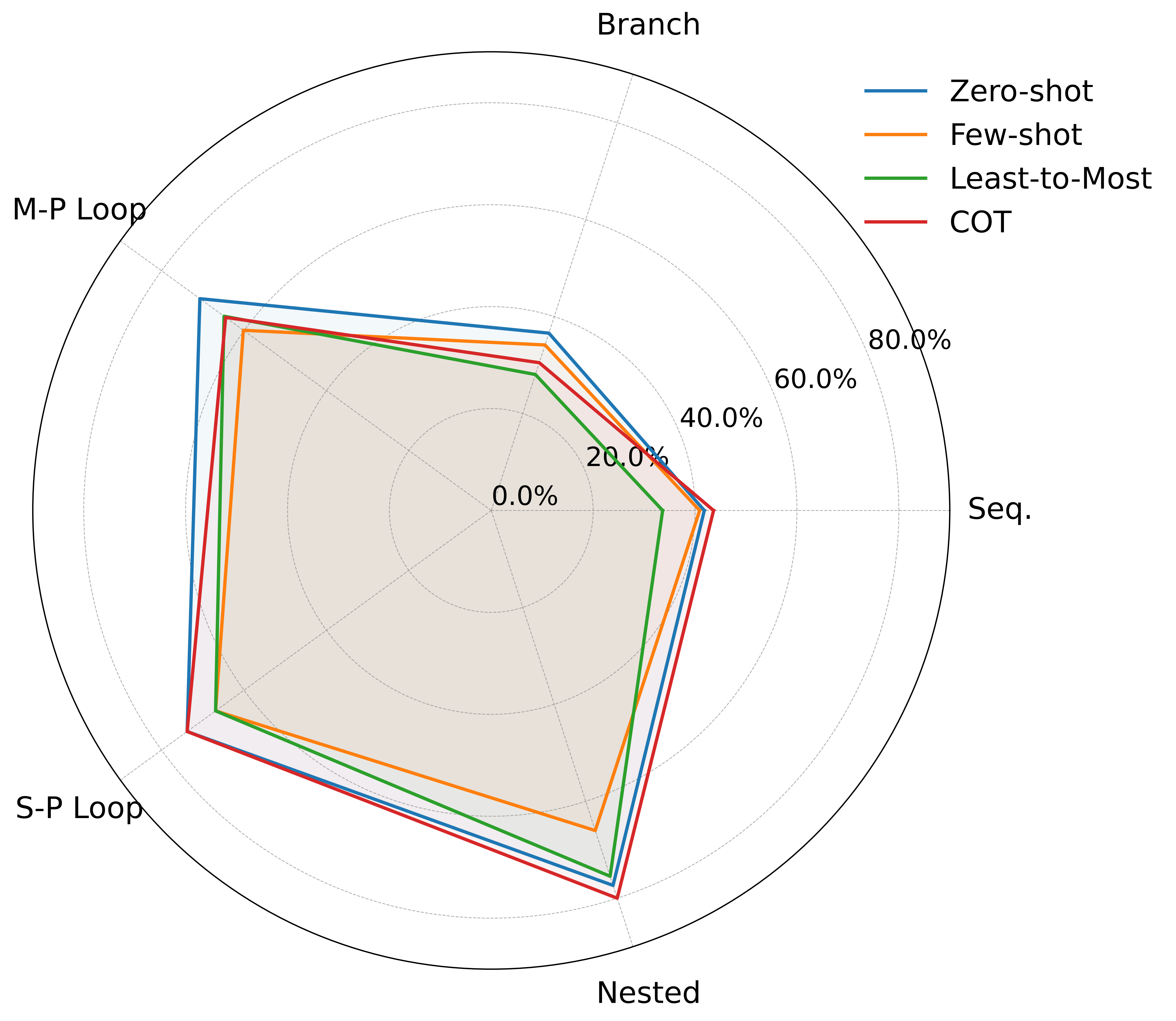}
    \caption{Verification Failures of GPT-4o}
    \label{fig:enter-label}
\end{figure}

\begin{figure}[h]
    \centering
    \includegraphics[width=0.85\linewidth]{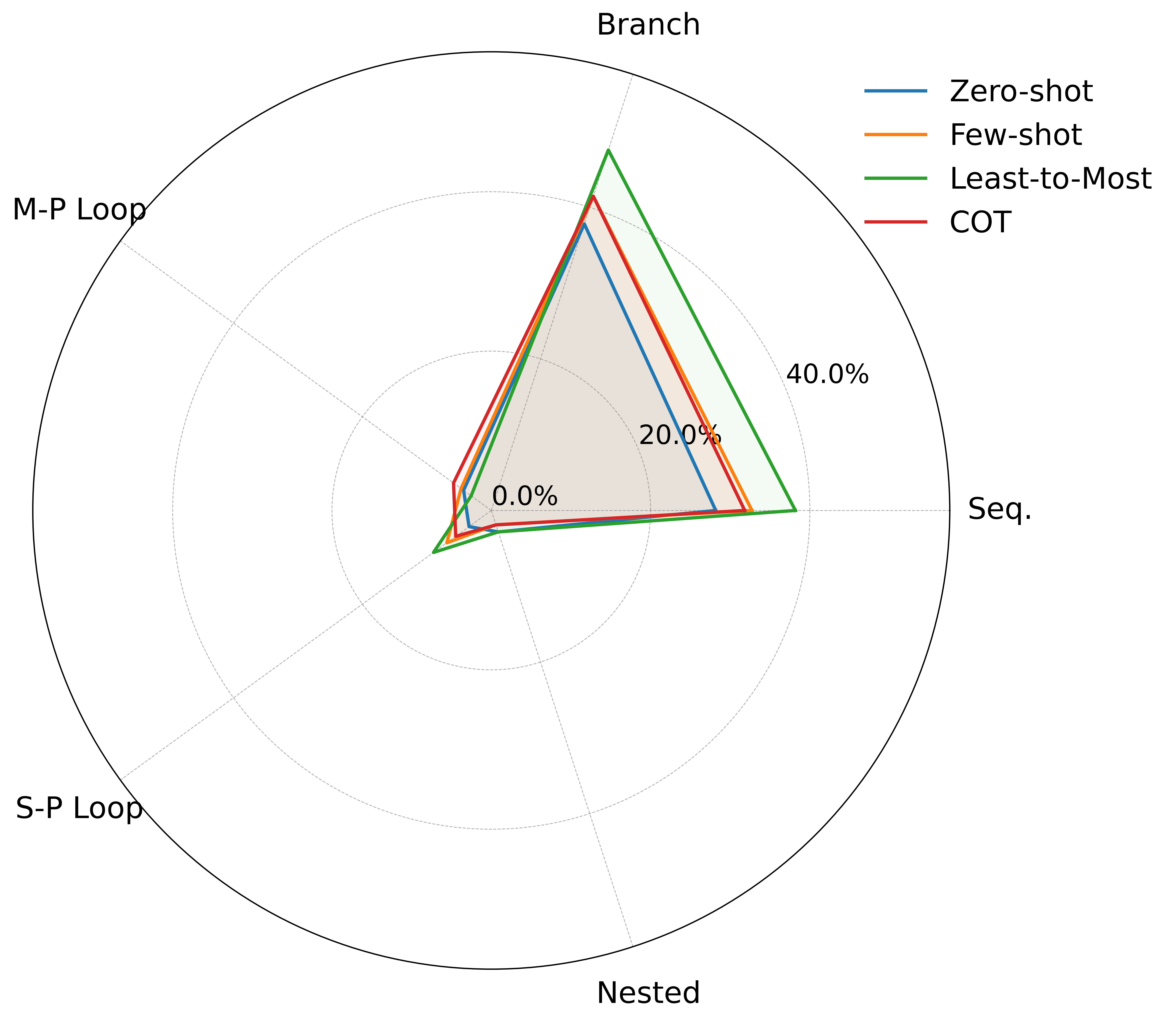}
    \caption{Verification Successes of GPT-4o}
    \label{fig:enter-label}
\end{figure}

\begin{figure}[h]
    \centering
    \includegraphics[width=0.85\linewidth]{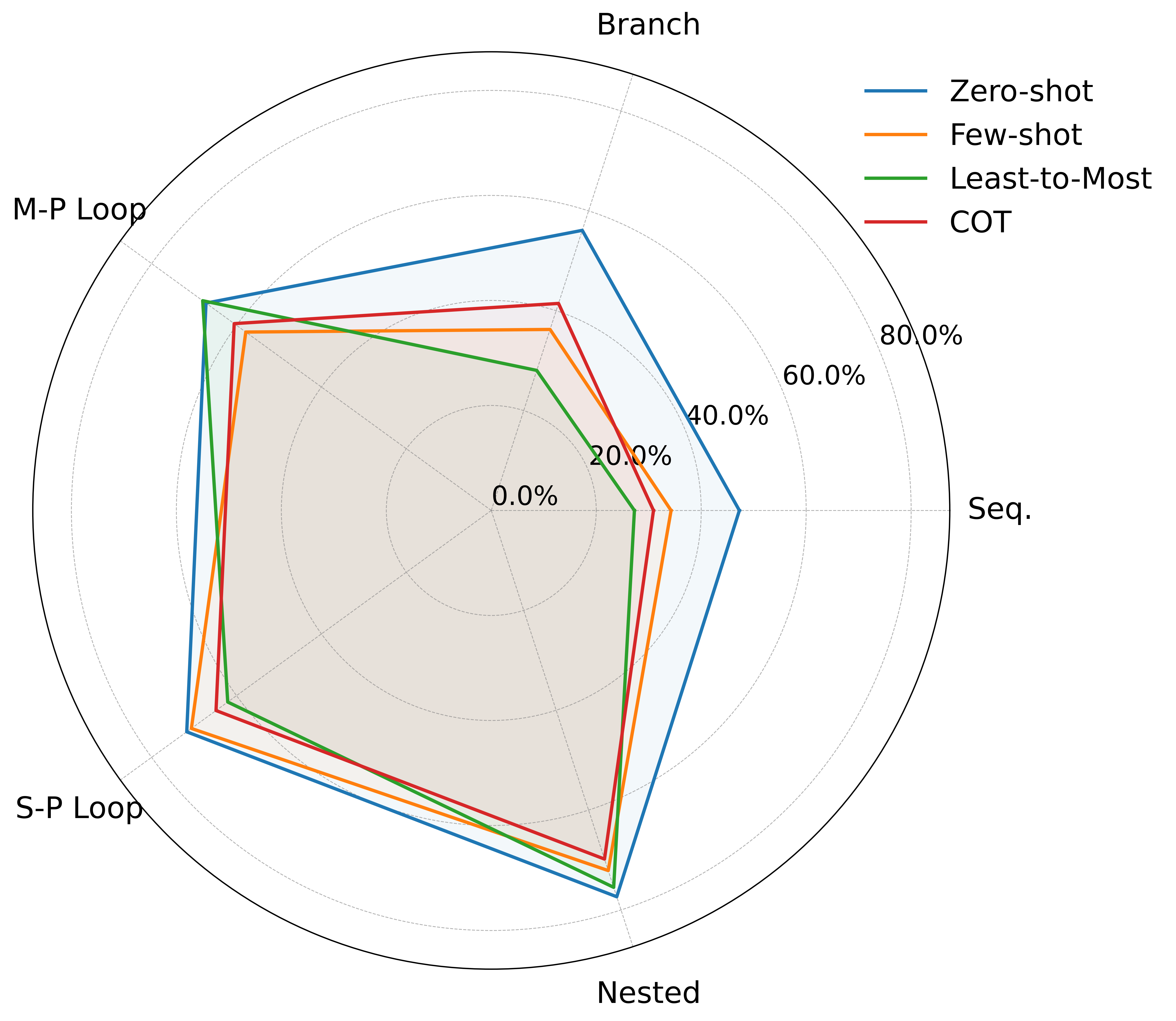}
    \caption{Verification Failures of DeepSeek-V3}
    \label{fig:enter-label}
\end{figure}

\begin{figure}[h]
    \centering
    \includegraphics[width=0.85\linewidth]{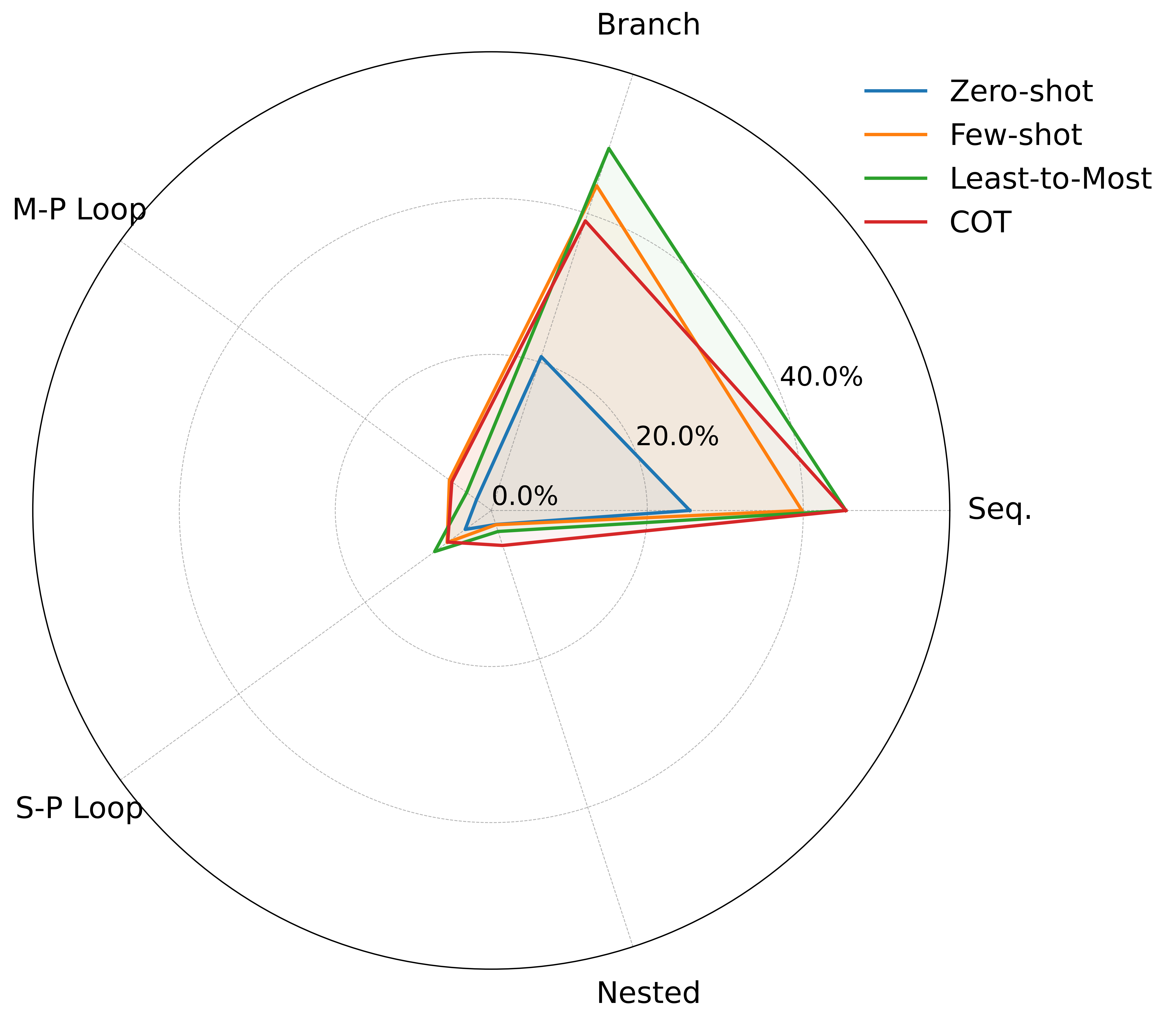}
    \caption{Verification Successes of DeepSeek-V3}
    \label{fig:enter-label}
\end{figure}

\section{Verifiable LLM-generated Specifications}

In this section, we provide several successful cases of LLM-generated specification, which can be verified to be correct by OpenJML.

\begin{lstlisting}[aboveskip=2pt, belowskip=2pt, basicstyle=\scriptsize\ttfamily]
class Maximum {
    
    /*@
    @ requires a >= 0 && b >= 0;
    @ ensures \result == a || \result == b;
    @ ensures \result >= a && \result >= b;
    @*/
    public static int maximum(int a, int b) {
        return a > b? a : b;
    }
}
\end{lstlisting}

\begin{lstlisting}[aboveskip=2pt, belowskip=2pt, basicstyle=\scriptsize\ttfamily]
class Lcopy {
    
    //@ requires xs!= null && xs.length > 0;
    //@ ensures \result!= null && \result.length == xs.length;
    //@ ensures Arrays.equals(\result, xs);
    public static int[] lcopy(int[] xs) {
        int[] res = new int[xs.length];
        System.arraycopy(xs, 0, res, 0, xs.length);
        return res;
    }
}
\end{lstlisting}

\begin{lstlisting}[aboveskip=2pt, belowskip=2pt, basicstyle=\scriptsize\ttfamily]

\end{lstlisting}

\begin{lstlisting}[aboveskip=2pt, belowskip=2pt, basicstyle=\scriptsize\ttfamily]
public class CountCharac {

    //@ requires str1!= null;
    //@ ensures \result == str1.length();
    public static int countCharac(String str1) {
        return str1.length();
    }
}
\end{lstlisting}

\begin{lstlisting}[aboveskip=2pt, belowskip=2pt, basicstyle=\scriptsize\ttfamily]
import java.io.*;
import java.lang.*;
import java.util.*;
import java.math.*;

class AsciiValue {
    //@ requires k!= null && k.length() == 1;
    //@ ensures \result == (k.length() == 1? (int) k.charAt(0) : -1);
    public static int asciiValue(String k) {
        if (k.length() == 1) {
            return (int) k.charAt(0);
        } else {
            return -1;
        }
    }
}
\end{lstlisting}

\begin{lstlisting}[aboveskip=2pt, belowskip=2pt, basicstyle=\scriptsize\ttfamily]
class CheckOddParity {
    
    //@ requires x <= Integer.MAX_VALUE && x >= Integer.MIN_VALUE;
    //@ ensures \result == (x % 2!= 0);
    public static Boolean checkOddParity(int x) {
        return x % 2!= 0;
    }
}
\end{lstlisting}
\begin{lstlisting}[aboveskip=2pt, belowskip=2pt, basicstyle=\scriptsize\ttfamily]
class SwapArray {
    
    //@ requires newArray!= null && 0 <= newArray.length < Integer.MAX_VALUE;
    //@ ensures \old(newArray.length) == 0 ==> newArray.length == 0;
    //@ ensures \old(newArray.length) > 0 ==> newArray[0] == \old(newArray[\old(newArray.length)-1]);
    //@ ensures \old(newArray.length) > 0 ==> newArray[\old(newArray.length)-1] == \old(newArray[0]);
    public static int[] swapArray(int[] newArray) {
        if (newArray.length == 0) {
            return newArray;
        }
        int first = newArray[0];
        int last = newArray[newArray.length - 1];
        newArray[0] = last;
        newArray[newArray.length - 1] = first;
        return newArray;
    }
}
\end{lstlisting}

\section{Failure Analysis}
\label{appx:success_analysis}

Since each specification can contain failures at multiple locations, our analysis begins by separating these errors into atomic errors. We then conduct a manual analysis of each error to identify common patterns in their error messages. For example, failures related to postconditions consistently include the following string in their messages: "The prover cannot establish an assertion (Postcondition)". Based on these patterns, we build a simple pattern-matching technique to classify failures. This process was applied iteratively until all unidentified patterns were resolved. In total, we identified XXX distinct types of failures. The ten most frequently occurring failure types are presented and discussed in detail.

\textbf{Syntax Errors.}. LLMs frequently produce formal specifications containing syntactic errors. Although the incidence of such issues can be partially mitigated through the provision of more detailed prompts or instructions, they remain a common occurrence. For instance, in the following code snippet, GPT-4o erroneously includes the `assignable` keyword within a loop invariant clause, thereby violating the syntactic rules of the Java Modeling Language.

\begin{lstlisting}[aboveskip=2pt, belowskip=2pt, basicstyle=\scriptsize\ttfamily]
class ReverseArrayLists {
        // Some code here
        
        /*@
          @ assignable result[*];
          @*/
        for (int i = 0; i < lists.length; i++) {
            // Some code here
        }
        
        return result;
    }
}
\end{lstlisting}

\textbf{Unsupported Inductive Quantifiers}. LLMs frequently necessitate the use of inductive quantifiers, such as `num\_of`, `sum`, and `product`, within formal specifications, as in the following example:

\begin{lstlisting}[aboveskip=2pt, belowskip=2pt, basicstyle=\scriptsize\ttfamily]
import java.io.*;
import java.lang.*;
import java.util.*;
import java.math.*;

class CountSetBits {
    
    /*@ 
      @ requires n >= 0;
      @ ensures \result >= 0;
      @ ensures \result == (\sum int i; 1 <= i && i <= n; Integer.bitCount(i));
      @*/
    public static int countSetBits(int n) {
        int count = 0;
        
        //@ maintaining count >= 0;
        //@ maintaining count == (\sum int j; 1 <= j && j <= i-1; Integer.bitCount(j));
        //@ maintaining i >= 1 && i <= n+1;
        for (int i = 1; i <= n; i++) {
            count += Integer.bitCount(i);
        }
        return count;
    }
}
\end{lstlisting}

However, these constructs are generally unsupported by deductive verification techniques. The semantic interpretation of such operators inherently demands inductive reasoning, encompassing tasks like counting elements, aggregating values over a range, or computing multiplicative products. These forms of reasoning pose significant challenges for Satisfiability Modulo Theories (SMT) solvers, which underpin tools such as OpenJML and the majority of deductive verification frameworks. Consequently, to circumvent the limitations associated with the direct use of inductive quantifiers, practitioners in formal methods must reformulate specifications using inductive steps. This involves the definition of auxiliary mathematical functions and lemmas via model methods, thereby facilitating the verifier's ability to reason about the intended program behavior.

\textbf{Postcondition Failures}. This error occurs when the postcondition—a condition that must hold true after the execution of a program or function—is not satisfied. This type of semantic error typically arises when verifiers are unable to confirm that the program’s logic guarantees the postcondition under all valid inputs and scenarios. The root causes of this error include: (1) an incorrect or incomplete postcondition, (2) wrong/weak preconditions that prevent the program from reaching a state where the postcondition holds, or (3) incomplete reasoning about the programs, leading to insufficient information for the verifier to verify the postcondition. An example of this class of error is provided below.

\begin{lstlisting}[aboveskip=2pt, belowskip=2pt, basicstyle=\scriptsize\ttfamily]
import java.io.*;
import java.lang.*;
import java.util.*;
import java.math.*;

class ReArrangeTuples {

    /*@ 
      @ requires testArray != null && ordArray != null;
      @ requires testArray.length == ordArray.length;
      @ requires (\forall int i; 0 <= i < testArray.length; testArray[i] != null && testArray[i].length == 2);
      @ ensures \result != null && \result.length == ordArray.length;
      @ ensures (\forall int i; 0 <= i < ordArray.length; 
      @             (\exists int j; 0 <= j < testArray.length; 
      @                 \result[i][0] == ordArray[i] && \result[i][0] == testArray[j][0] && \result[i][1] == testArray[j][1]));
      @*/
    public static int[][] reArrangeTuples(int[][] testArray, int[] ordArray) {
        int n = ordArray.length;
        int[][] res = new int[n][2]; 

        int index = 0;

        /*@ 
          @ loop_invariant 0 <= i <= n;
          @ loop_invariant 0 <= index <= i;
          @ loop_invariant (\forall int k; 0 <= k < index; 
          @     (\exists int j; 0 <= j < testArray.length; 
          @         res[k][0] == ordArray[k] && res[k][0] == testArray[j][0] && res[k][1] == testArray[j][1]));
          @ decreases n - i;
          @*/
        for (int i = 0; i < n; i++) {
            /*@ 
              @ loop_invariant 0 <= j <= testArray.length;
              @ loop_invariant 0 <= index <= i;
              @ loop_invariant (\forall int k; 0 <= k < index; 
              @     (\exists int m; 0 <= m < testArray.length; 
              @         res[k][0] == ordArray[k] && res[k][0] == testArray[m][0] && res[k][1] == testArray[m][1]));
              @ decreases testArray.length - j;
              @*/
            for (int j = 0; j < testArray.length; j++) {
                if (testArray[j][0] == ordArray[i]) {
                    res[index++] = testArray[j];
                    break;
                }
            }
        }

        return res;
    }
}

// Error Message:
// /tmp/ReArrangeTuples.java:49: verify: The prover cannot establish an assertion (Postcondition: /tmp/ReArrangeTuples.java:14:) in method reArrangeTuples
\end{lstlisting}

\textbf{Arithmetic Operation Range Failures.}. This error occurs when arithmetic overflows cause computations to exceed the allowable range of values. Such errors frequently occur because LLMs often fail to adequately reason about the necessary bounds of variables within a program. As a result, they may generate specifications or code that do not account for the potential of arithmetic overflow under certain input conditions. An example of this class of error is provided below.

\begin{lstlisting}[aboveskip=2pt, belowskip=2pt, basicstyle=\scriptsize\ttfamily]
// failed

import java.io.*;
import java.lang.*;
import java.util.Arrays;

class RemoveNested {
    
    /*@ 
      @ requires testTup != null;
      @ requires (\forall int i; 0 <= i && i < testTup.length;
      @          testTup[i] == null ||
      @          testTup[i] instanceof Integer ||
      @          (testTup[i] instanceof Object[] &&
      @           (\forall int j; 0 <= j && j < ((Object[])testTup[i]).length;
      @            ((Object[])testTup[i])[j] == null ||
      @            ((Object[])testTup[i])[j] instanceof Integer)));
      @ ensures \result != null;
      @ ensures \result.length <= testTup.length;
      @ ensures (\forall int i; 0 <= i && i < \result.length; 
      @         (\exists int j; 0 <= j && j < testTup.length;
      @          testTup[j] instanceof Integer && \result[i] == (Integer)testTup[j]));
      @*/
    public static int[] removeNested(Object[] testTup) {
        int[] temp = new int[testTup.length];
        int count = 0;
        
        //@ maintaining 0 <= count && count <= testTup.length;
        //@ maintaining count <= temp.length;
        for (Object obj : testTup) {
            if (obj instanceof Object[]) {
                Object[] l = (Object[]) obj;
                for (Object e : l) {
                    if (e instanceof Integer) {
                    }
                }
            } else if (obj instanceof Integer) {
                temp[count++] = (Integer) obj;
            }
        }
        return Arrays.copyOf(temp, count);
    }
}

// Error Message:
// /tmp/RemoveNested.java:37: verify: The prover cannot establish an assertion (ArithmeticOperationRange) in method removeNested: overflow in int sum
\end{lstlisting}

\textbf{Loop Invariant Failures.}. This error occurs when the loop invariant, a condition that must hold true before the loop begins and remain true after each iteration, is not properly established or maintained. This semantic error typically arises when verifiers fail to confirm the correctness of the synthesized loop invariant. The causes of this error include: (1) an incorrect loop invariant, (2) wrong/weak preconditions that prevent the invariant from holding at the start of the loop, or (3) incomplete reasoning about the loop, leading to insufficient information for the verifier to verify the invariant. An example of this class of error is provided below.

\begin{lstlisting}[aboveskip=2pt, belowskip=2pt, basicstyle=\scriptsize\ttfamily]
// failed

import java.io.*;
import java.lang.*;
import java.util.*;
import java.math.*;

class GetGcd {

    //@ requires arr != null && arr.length > 0;
    //@ ensures (\forall int i; 0 <= i && i < arr.length; arr[i] >= 0);
    //@ ensures (\forall int i; 0 <= i && i < arr.length; \result <= arr[i]);
    //@ ensures (\exists int i; 0 <= i && i < arr.length; \result == arr[i]);
    public static int getGcd(int[] arr) {
        int result = 0;
        int min = arr[0];
        //@ maintaining 0 <= i && i <= arr.length;
        //@ maintaining (\forall int j; 0 <= j && j < i; arr[j] >= 0);
        //@ maintaining result <= min;
        //@ maintaining (\exists int j; 0 <= j && j < i; result == arr[j]);
        for (int i = 1; i < arr.length; i++) {
            if (arr[i] > min) {
                result = result > min ? result : min;
            } else {
                min = arr[i];
            }
        }
        return result;
    }
}

// Error Message:
// /tmp/NthNums.java:24: verify: The prover cannot establish an assertion (Postcondition: /tmp/NthNums.java:13:) in method nthNums

\end{lstlisting}

\textbf{Assertion Failures.}. This error occurs when an assertion, a condition that must hold true at a specific point in the program, evaluates to false during execution. This type of error typically arises due to (1) incorrect assertions, (2) incomplete reasoning about program behavior, leading to insufficient information for the verifier to verify the assertions, or (3) insufficient preconditions that fail to guarantee the assertion. An example of this class of error is provided below.

\begin{lstlisting}[aboveskip=2pt, belowskip=2pt, basicstyle=\scriptsize\ttfamily]
// failed

import java.io.*;
import java.lang.*;
import java.util.*;
import java.math.*;

class FindMinDiff {

    //@ requires arr != null;
    //@ requires n > 0;
    //@ requires n == arr.length;
    //@ ensures \result >= 0;
    //@ ensures (\forall int i, j; 0 <= i && i < n && 0 <= j && j < n; Math.abs(arr[i] - arr[j]) >= \result);
    public static int findMinDiff(int[] arr, int n) {
        //@ assume n > 0 && arr != null && n == arr.length;

        int minDiff = Integer.MAX_VALUE;
        //@ maintaining minDiff == Integer.MAX_VALUE || (\forall int k; 0 <= k && k < i; minDiff <= Math.abs(arr[i] - arr[k]));
        for (int i = 0; i < n - 1; i++) {
            //@ maintaining minDiff == Integer.MAX_VALUE || (\forall int k; 0 <= k && k < i; minDiff <= Math.abs(arr[i] - arr[k]));
            for (int j = i + 1; j < n; j++) {
                int diff = Math.abs(arr[i] - arr[j]);
                //@ assert diff == Math.abs(arr[i] - arr[j]);

                if (diff < minDiff) {
                    minDiff = diff;
                }
                //@ assert minDiff <= diff;
            }
        }
        //@ assert (\forall int i, j; 0 <= i && i < n && 0 <= j && j < n; Math.abs(arr[i] - arr[j]) >= minDiff);
        return minDiff;
    }
}

// Error Message:
// /tmp/FindMinDiff.java:31: verify: The prover cannot establish an assertion (Assert) in method findMinDiff

\end{lstlisting}

\textbf{Null Dereference.}. This error occurs when a null pointer is dereferenced, leading to undefined behavior or runtime failures. These issues typically arise since LLMs do not properly reason about preconditions for ensuring the non-nullness of objects, such as arrays. An example of this class of error is provided below.

\begin{lstlisting}[aboveskip=2pt, belowskip=2pt, basicstyle=\scriptsize\ttfamily]
// failed

import java.io.*;
import java.lang.*;
import java.util.*;
import java.math.*;

class FindCharLong {
    
    /**
     * Finds all words in the given text that are at least 4 characters long.
     * 
     * @param text the input string to search for words
     * @return an array of words with at least 4 characters
     * @throws NullPointerException if text is null
     * 
     * @ensures \result != null && (\forall int i; 0 <= i && i < \result.length; \result[i].length() >= 4) &&
     *          (\forall String word; (\exists int i; 0 <= i && i < \result.length; \result[i].equals(word)) <==> 
     *          (word.length() >= 4 && (\exists int j; 0 <= j && j < words.length; words[j].equals(word))))
     */
    public static String[] findCharLong(String text) {
        //@ requires text != null;
        //@ ensures \result != null;
        //@ ensures (\forall int i; 0 <= i && i < \result.length; \result[i].length() >= 4);

        String[] words = text.split("\\s+");
        int count = 0;

        // Count words with 4 or more characters
        /*@ loop_invariant 0 <= count && count <= words.length;
          @ loop_invariant (\forall int k; 0 <= k && k < count; words[k].length() >= 4);
          @ decreases words.length - count;
          @*/
        for (String word : words) {
            if (word.length() >= 4) {
                count++;
            }
        }

        String[] result = new String[count];
        int index = 0;

        // Collect words with 4 or more characters
        /*@ loop_invariant 0 <= index && index <= count;
          @ loop_invariant (\forall int k; 0 <= k && k < index; result[k].length() >= 4);
          @ loop_invariant (\forall int k; 0 <= k && k < index; 
          @     (\exists int j; 0 <= j && j < words.length; result[k].equals(words[j])));
          @ decreases words.length - index;
          @*/
        for (String word : words) {
            if (word.length() >= 4) {
                result[index++] = word;
            }
        }
        return result;
    }
}

// Error Message:
// /tmp/FindCharLong.java:25: verify: The prover cannot establish an assertion (PossiblyNullDeReference) in method findCharLong


\end{lstlisting}

\textbf{Divide By Zero.}. This error occurs when a division operation attempts to divide by zero, leading to undefined behavior. These issues typically arise due to missing or incomplete specifications of the values of variables. The root causes can vary and include insufficient preconditions, missing loop invariants, or the absence of assertions to enforce non-zero denominators. An example of this class of error is provided below.

\begin{lstlisting}[aboveskip=2pt, belowskip=2pt, basicstyle=\scriptsize\ttfamily]
import java.io.*;
import java.lang.*;
import java.util.*;
import java.math.*;

class RoundNum {
    
    /**
     * @requires m > 0; // The divisor must be positive.
     * @ensures \result >= 0; // The result will be non-negative.
     * @ensures \result % m == 0; // The result will be a multiple of m.
     * @ensures \result <= n; // The result will be less than or equal to n.
     * @ensures (\forall int i; \result < i && i <= n; i % m != 0); 
     *   // There is no multiple of m between \result and n.
     */
    public static int roundNum(int n, int m) {
        /*@ loop_invariant 
              0 <= i <= n && 
              (\forall int j; i < j && j <= n; j % m != 0);
            // i is non-negative, i is less than or equal to n, and
            // there is no multiple of m between i and n.
          @*/
        for (int i = n; i >= 0; i--) {
            //@ assert i >= 0; // i is non-negative.
            if (i % m == 0) {
                return i;
            }
        }
        //@ assert false; // This code should never be reached.
        return -1;
    }
}

// Error Message:
// /tmp/RoundNum.java:26: verify: The prover cannot establish an assertion (PossiblyDivideByZero) in method roundNum

\end{lstlisting}

\textbf{Invalid Specifications.}. LLMs, particularly open-source variants, frequently produce specifications expressed in natural language rather than in the required formal syntax. Such outputs are not interpretable by formal verification tools and thus constitute invalid specifications. 

\textbf{Array Index Failures.}. This error occurs when an array index exceeds its valid bounds, leading to potential runtime failures. These issues typically arise due to missing conditions or specifications regarding the array's length or the bounds of the index. An illustrative example of this type of error is presented below.

\begin{lstlisting}[aboveskip=2pt, belowskip=2pt, basicstyle=\scriptsize\ttfamily]
// failed

import java.io.*;
import java.lang.*;
import java.util.*;
import java.math.*;

class Sum {
    /*@ public normal_behavior
      @ requires n >= 0;
      @ requires n < Integer.MAX_VALUE;
      @ ensures \result >= 0;
      @ ensures n <= 1 ==> \result == 0;
      @*/
    public static int sum(int n) {
        //@ assume n >= 0;
        int[] sumOfPrimeDivisors = new int[n + 1];
        
        //@ assume sumOfPrimeDivisors != null;
        //@ assume sumOfPrimeDivisors.length == n + 1;
        sumOfPrimeDivisors[0] = 0;
        sumOfPrimeDivisors[1] = 0;

        /*@ loop_invariant 2 <= i && i <= n + 1;
          @ loop_invariant \forall int k; 0 <= k && k < i; 
          @     sumOfPrimeDivisors[k] >= 0;
          @ loop_invariant sumOfPrimeDivisors[0] == 0;
          @ loop_invariant sumOfPrimeDivisors[1] == 0;
          @ decreases n - i;
          @*/
        for (int i = 2; i <= n; i++) {
            if (sumOfPrimeDivisors[i] == 0) {
                /*@ loop_invariant j >= i;
                  @ loop_invariant j <= n + i;
                  @ loop_invariant \forall int k; i <= k && k < j && k % i == 0; 
                  @     sumOfPrimeDivisors[k] >= i;
                  @ decreases n - j;
                  @*/
                for (int j = i; j <= n; j += i) {
                    //@ assume sumOfPrimeDivisors[j] + i <= Integer.MAX_VALUE;
                    sumOfPrimeDivisors[j] += i;
                }
            }
        }

        //@ assert sumOfPrimeDivisors[n] >= 0;
        return sumOfPrimeDivisors[n];
    }
}

// Error Message:
// /tmp/Sum.java:21: verify: The prover cannot establish an assertion (PossiblyTooLargeIndex) in method sum

\end{lstlisting}

\section{Prompts}
\label{appx:prompts}

In this section, we present prompt templates used in this study for specification generation, including zero-shot, few-shot, chain-of-thought, and least-to-most prompts.

\begin{llmpromptbox}{Zero-shot prompt}
(System) You are an expert in Java Modeling Language (JML). You will be provided with Java code snippets and their task descriptions. Your task is to generate JML specifications for the given Java code. The specifications should be written as annotations within the Java code and must be compatible with the OpenJML tool for verification. Ensure the specifications include detailed preconditions, postconditions, necessary loop invariants, invariants, assertions, and any relevant assumptions.

(User) Please generate JML specifications for the provided Java code.

\#\#\# CODE

\{code\}

\end{llmpromptbox}

\begin{llmpromptbox}{Few-shot prompt}
(System) You are an expert in Java Modeling Language (JML). 
You will be provided with Java code snippets. 
Your task is to generate JML specifications for the given Java code. 
The specifications should be written as annotations within the Java code and must be compatible with the OpenJML tool for verification. 
Ensure the specifications include detailed preconditions, postconditions, necessary loop invariants, invariants, assertions, and any relevant assumptions.

Please also adhere to the following syntax guidelines for JML:

JML text is written in comments that either:

a) begin with //@ and end with the end of the line, or

b) begin with /*@ and end with */. Lines within such a block comment may 
have the first non-whitespace characters be a series of @ symbols.

\{examples\}

(User) Please generate JML specifications for the provided Java code.

\#\#\# CODE

\{code\}
\end{llmpromptbox}

\begin{llmpromptbox}{Chain-of-thought prompt}
(System) You are an expert in Java Modeling Language (JML). 
You will be provided with Java code snippets. 
Your task is to generate JML specifications for the given Java code. 
The specifications should be written as annotations within the Java code and must be compatible with the OpenJML tool for verification. 
Ensure the specifications include detailed preconditions, postconditions, necessary loop invariants, invariants, assertions, and any relevant assumptions.

Please also adhere to the following syntax guidelines for JML:

JML text is written in comments that either:

a) begin with //@ and end with the end of the line, or

b) begin with /*@ and end with */. Lines within such a block comment may 
have the first non-whitespace characters be a series of @ symbols.

\{examples\}

(User) Please generate JML specifications for the provided Java code.

\#\#\# CODE

\{code\}

Let's think step by step!
\end{llmpromptbox}

\begin{llmpromptbox}{Least-to-Most prompt}
(System) You are an expert in Java Modeling Language (JML). 
You will be provided with Java code snippets. 
Your task is to generate JML specifications for the given Java code. 
The specifications should be written as annotations within the Java code and must be compatible with the OpenJML tool for verification. 
Ensure the specifications include detailed preconditions, postconditions, necessary loop invariants, invariants, assertions, and any relevant assumptions.

Please also adhere to the following syntax guidelines for JML:

JML text is written in comments that either:

a) begin with //@ and end with the end of the line, or

b) begin with /*@ and end with */. Lines within such a block comment may 
have the first non-whitespace characters be a series of @ symbols.

\{examples\}

(User) Please generate JML specifications for the provided Java code.

\#\#\# CODE

\{code\}

Let's break down this problem:

1. What are the weakest preconditions for the code? Be sure to include preconditions related to nullness and arithmetic bounds.

2. What are the strongest postconditions for the code?

3. What necessary specifications are required to prove the above post-conditions? This includes loop invariants, assertions, assumptions, and ranking functions.

After answering these questions, let's generate the specifications for the code and provide solution after `\#\#\# SPECIFCIATION'

\end{llmpromptbox}

\section{Self-Repair Prompts}
\label{appx:repair_prompts}

In this section, we illustrate several repair prompts designed to address the most common errors. For a complete list of all repair prompts, please refer to our repository.

\begin{llmpromptbox}{Fixing prompt for Syntax Errors}
(System) You are an experts on Java Modeling Language (JML). Your task is to fix the JML specifications annotated in the target Java code. You will be provided the error messages from the OpenJML tool and you need to fix the specifications accordingly.

(User) The following Java code is annotated with JML specifications:

\{current specification\}

OpenJML Verification tool failed to verify the specifications given above, with error information as follows:

\#\#\# ERROR MESSAGE:

\{error messages\}

\#\#\# ERROR TYPES: Syntax Error

To resolve the syntax error, you should consider the following steps:

1. Identify whether the error is due to a Java syntax issue or a JML syntax issue.

2. Review the code to identify the specific location and nature of the syntax error.

3. Correct the syntax error based on the language rules and conventions.

Please refine the specifications so that they can pass verification. Provide the specifications for the code and include the solution written between triple backticks, after `\#\#\# FIXED SPECIFICATION`.
\end{llmpromptbox}

\begin{llmpromptbox}{Fixing prompt for Unsupported Sum/NumOf/Product Quantifier Expressions}
(System) You are an experts on Java Modeling Language (JML). Your task is to fix the JML specifications annotated in the target Java code. You will be provided the error messages from the OpenJML tool and you need to fix the specifications accordingly.

(User) The following Java code is annotated with JML specifications:

\{current specification\}

OpenJML Verification tool failed to verify the specifications given above, with error information as follows:

\#\#\# ERROR MESSAGE:

\{error messages\}

\#\#\# ERROR TYPES: Unsupported Sum/NumOf/Product Quantifier Expressions

OpenJML does not fully support JML's inductive quantifiers like \textbackslash num\_of, \textbackslash sum, and \textbackslash product in specifications. These operators require inductive reasoning (e.g., counting elements, summing values over a range, or computing products), which is difficult for SMT solvers (the engines behind OpenJML and most of deductive verification tools) to handle.

To avoid the use of \textbackslash sum, \textbackslash num\_of, and \textbackslash product quantifiers in your JML specifications, you can express your specifications using induction steps to help OpenJML's verifiers to reason about your code. You can do this by define mathematical functions and lemmas through model methods. 
For example, you can should not use \textbackslash product quantifier in the following specifications:

\{Examples with reasoning\}

Please refine the specifications so that they can pass verification. Provide the specifications for the code and include the solution written between triple backticks, after `\#\#\# FIXED SPECIFICATION`.
\end{llmpromptbox}

\begin{llmpromptbox}{Fixing prompt for Unsupported Min/Max Quantifier Expressions}
(System) You are an experts on Java Modeling Language (JML). Your task is to fix the JML specifications annotated in the target Java code. You will be provided the error messages from the OpenJML tool and you need to fix the specifications accordingly.

(User) The following Java code is annotated with JML specifications:

\{current specification\}

OpenJML Verification tool failed to verify the specifications given above, with error information as follows:

\#\#\# ERROR MESSAGE:

\{error messages\}

\#\#\# ERROR TYPES: Unsupported Min/Max Quantifier Expressions

OpenJML does not fully support JML's inductive quantifiers like \textbackslash min, \textbackslash max in specifications. These operators require inductive reasonings, which is difficult for SMT solvers (the engines behind OpenJML and most of deductive verification tools) to handle.

To avoid the use of \textbackslash min and \textbackslash max quantifiers in your JML specifications, you can use the \textbackslash forall quantifier to express your specifications. 
For example, you should not use \textbackslash max quantifier in the following specifications:

\{Examples with reasoning\}

Please refine the specifications so that they can pass verification. Provide the specifications for the code and include the solution written between triple backticks, after `\#\#\# FIXED SPECIFICATION`.
\end{llmpromptbox}

\begin{llmpromptbox}{Fixing prompt for Loop Invariant Failures}
(System) You are an experts on Java Modeling Language (JML). Your task is to fix the JML specifications annotated in the target Java code. You will be provided the error messages from the OpenJML tool and you need to fix the specifications accordingly.

(User) The following Java code is annotated with JML specifications:

\{current specification\}

OpenJML Verification tool failed to verify the specifications given above, with error information as follows:

\#\#\# ERROR MESSAGE:

\{error messages\}

\#\#\# ERROR TYPES: Loop Invariant Failures

This error occurs when the loop invariant, a condition that must hold true before the loop begins and remain true after each iteration, is not properly established or maintained. This semantic error typically arises when verifiers fail to confirm the correctness of the synthesized loop invariant. The causes of this error include: (1) an incorrect loop invariant, (2) wrong/weak preconditions that prevent the invariant from holding at the start of the loop, or (3) incomplete reasoning about the loop, leading to insufficient information for the verifier to verify the invariant.

To resolve the error, please consider the following steps:

1. Carefully review the loop invariant to ensure it correctly captures the necessary conditions that hold true before and after each iteration of the loop.

2. Carefully examine preconditions to ensure they are strong enough to establish the loop invariant at the beginning of the loop.

3. Add additional assertions or assumptions within the loop to help the verifier reason about the loop invariant.

For example, consider the following code snippet with a loop invariant failure:

\{Examples with reasoning\}

Please refine the specifications so that they can pass verification. Provide the specifications for the code and include the solution written between triple backticks, after `\#\#\# FIXED SPECIFICATION`.
\end{llmpromptbox}

\begin{llmpromptbox}{Fixing prompt for Post-Condition Failures}
(System) You are an experts on Java Modeling Language (JML). Your task is to fix the JML specifications annotated in the target Java code. You will be provided the error messages from the OpenJML tool and you need to fix the specifications accordingly.

(User) The following Java code is annotated with JML specifications:

\{current specification\}

OpenJML Verification tool failed to verify the specifications given above, with error information as follows:

\#\#\# ERROR MESSAGE:

\{error messages\}

\#\#\# ERROR TYPES: Post-condition Failures

This error occurs when the postcondition, a condition that must hold true after the execution of a program or function, is not satisfied. This type of semantic error typically arises when verifiers are unable to confirm that the program’s logic guarantees the postcondition under all valid inputs and scenarios. The causes of this error include: (1) an incorrect or incomplete postcondition, (2) wrong/weak preconditions that prevent the program from reaching a state where the postcondition holds, or (3) incomplete reasoning about the programs, leading to insufficient information for the verifier to verify the postcondition.

To resolve the error, please consider the following steps:

1. Review the postcondition to ensure it correctly captures the expected behavior of the program or function.

2. Check the preconditions to ensure they are strong enough to reach a state where the postcondition holds.

3. Add additional assertions or assumptions within the program or function to help the verifier reason about the postcondition.

For example, consider the following code snippet with a postcondition failure:

\{Examples with reasoning\}

Please refine the specifications so that they can pass verification. Provide the specifications for the code and include the solution written between triple backticks, after `\#\#\# FIXED SPECIFICATION`.
\end{llmpromptbox}

\end{document}